\documentclass{pasj00}

\begin{document}
\SetRunningHead{Bautz et al.}{Suzaku Observations of Abell 1795}
\Received{}
\Accepted{}

\title{Suzaku Observations of Abell 1795:\\
Cluster Emission to R$_{\bf 200}$}

\author{Marshall W. \textsc{Bautz}, Eric D. \textsc{Miller}}
\affil{Kavli Institute for Astrophysics \& Space Research, Massachusetts Institute of Technology, \\ Cambridge, MA USA}
\email{mwb@space.mit.edu}
\author{Jeremy S. \textsc{Sanders}}
\affil{Cambridge University, Cambridge, UK}
\author{Keith A. \textsc{Arnaud}, Richard F. \textsc{Mushotzky}, F. Scott \textsc{Porter}}
\affil{NASA's Goddard Space Flight Center, Greenbelt, MD USA}
\author{Kiyoshi \textsc{Hayashida}}
\affil{Osaka University, Osaka, Japan}
\author{J. Patrick \textsc{Henry}}
\affil{University of Hawaii, Honolulu, HI USA}
\author{John P. \textsc{Hughes}}
\affil{Rutgers University, New Brunswick, NJ USA}
\author{Madoka \textsc{Kawaharada}, Kazuo \textsc{Makashima}, Mitsuhiro \textsc{Sato}} 
\affil{University of Tokyo, Tokyo, Japan}
\and
\author{Takayuki \textsc{Tamura}}
\affil{Institute of Space and Astronautical Sciences, 
Japan Aerospace Exploration Agency, \\Sagamihara, Japan}


\maketitle


\begin{abstract}
We report Suzaku observations of the galaxy cluster Abell 1795 that extend
to $r_{200} \approx 2$ Mpc, the radius within which the mean cluster mass
density is 200 times the cosmic critical density.  These observations are
the first to probe the state of the intracluster medium in this object at
$r > 1.3$ Mpc. We sample two disjoint sectors in  the cluster outskirts
($1.3 < r < 1.9$ Mpc) and detect X-ray emission in only one of them to a
limiting ($3\,\sigma$) soft X-ray surface brightness  of 
$B_{0.5-2~\rm{keV}} = 1.8 \times 10^{-12}$ erg s$^{-1}$ cm$^{-2}$ deg$^{-2}$, 
a level less than 20\% of the cosmic X-ray background brightness. 
We trace the run of temperature with radius at $r > 0.4$ Mpc and find that
it falls relatively rapidly ($T_{\rm{deprojected}} \propto r^{-0.9}$),
reaching a value about one third of its peak at the largest radius we
can measure it.  Assuming the intracluster medium is in hydrostatic
equilibrium and is polytropic, we find a polytropic index of
$\Gamma = 1.3^{+0.3}_{-0.2}$ and we estimate a mass of $4.1^{+0.5}_{-0.3}
\times 10^{14} M_{\odot}$ within 1.3 Mpc, somewhat ($2.7\,\sigma$) lower
than that reported by previous observers.  However, our observations
provide evidence for departure from hydrostatic equilibrium at radii as
small as $r \sim 1.3$ Mpc $\approx r_{500}$ in this apparently regular and
symmetrical cluster.
\end{abstract}

\section{Introduction}

Spatially resolved X-ray spectroscopy is essential for mapping the
distribution of mass, intracluster plasma, and heavy elements in galaxy
clusters.  Although Chandra and XMM-Newton have provided a
wealth of information on these quantities in the central regions of
clusters, relatively little is known about  their outer regions, at radii
greater than   $\sim 0.5 r_{200}$.  Here $r_{200}$ is the radius within
which the mean cluster density is 200 times the critical cosmic density. In 
the simplest case of the collapse of a spherical density perturbation in an 
Einstein-deSitter universe, it is expected that the dark and bayronic matter 
within $r_{200}$ will have reached dynamical equilibrium \citep{emn96}.  Cluster
emission from the vicinity of $r_{200}$ is difficult to observe efficiently
with Chandra and XMM-Newton because of their relatively high
(and, in the latter case, time-variable) levels of instrumental background.

The study of cluster outskirts is important for several reasons.  First, a
direct X-ray measurement of total cluster mass  (e.g., within $r_{200}$)
requires knowledge of the ICM temperature profile at this radius.  This is
especially important to facilitate comparison between X-ray and
weak-lensing mass profiles, as the latter are difficult to measure in the
inner regions of clusters. Second, structure and cluster formation models
make definite predictions about total mass and ICM distributions to these
radii \citep{nfw97,borg04,ron06}, so measurements in this radial range
offer a test of these models. Finally, to the extent that $r_{200}$ marks
the approximate boundary of the dynamically relaxed region of a cluster,
one might expect to find inhomogeneities in the ICM that could provide a
direct view of the accretion processes by which clusters grow. 

ROSAT has been used to characterize the X-ray surface brightness 
of the outer regions  ($r \sim r_{200}$) of individual  
clusters~\citep{vik99} and, using stacking techniques, of ensembles 
of objects~\citep{neu05}. Chandra surface brightness profiles 
in the range $r_{500} < r < r_{200}$ for a sample of relatively distant 
clusters have been studied by~\citet{ett08}.
Recent Suzaku observations have successfully probed out to and
beyond $r_{200}$ for a handful of clusters
\citep{geo08,Fujitaetal2008,Reiprichetal2008}.

In this paper we present Suzaku measurements of the surface
brightness and temperature of the intracluster medium of the X-ray luminous
($L_{X,bol} = 1.4 \times 10^{45}$ erg s$^{-1}$) cluster Abell 1795. This is
a relatively hot (spatially-averaged $kT = 5.3$ keV) 
cluster with a cool core \citep{ho01,vik06,sno08}.  Although Chandra
shows a 40-arcsec-long ($50\, h_{70}^{-1}$ kpc) plume ``trailing'' the
brightest cluster galaxy~\citep{fab01}, outside of the very center the
cluster is quite regular and appears relaxed.  From the observed
temperature, and scaling relations derived from
simulations~\citep{emn96,ea99} we estimate $r_{500} = 1.3\, h_{70}^{-1}$ Mpc,
consistent with the Chandra measurements of~\citet{vik06}, 
and $r_{200} = 1.9\, h_{70}^{-1}$ Mpc ($26\arcmin$) for our assumed
cosmology\footnote{We assume a flat, $\Lambda$CDM cosmology with 
$H_{0} = 70$ km s$^{-1}$ Mpc$^{-1}$, $\Omega_{M} = 0.3$ and 
$\Omega_{\Lambda} = 0.7$. One arc minute corresponds to  73 kpc at the
cluster redshift ($z=0.063$) in this cosmology. Unless otherwise noted,
quoted errors correspond to 90\% confidence intervals. }.  We exploit the relatively low and stable Suzaku
background to trace the temperature and ICM density to  the vicinity of
$r_{200}$. 

As  noted by \citet{bn98}, if  the
virial radius $r_{virial} = r_{\Delta_{v}}$ is derived from the collapse of a 
spherical top-hat density perturbation, 
assuming that the cluster has just virialized, 
then for an Einstein-deSitter cosmology 
$\Delta_{v} = 18 \pi^{2} \approx 180$, independent of redshift. 
In general, however, $\Delta_{v}$ is function of cosmology and redshift, 
and for  our $\Lambda$CDM cosmology at the redshift of 
Abell 1795 $\Delta_{v} \approx 100$, so 
the virial radius is considerably larger than $r_{200}$. 
For Abell 1795 we expect $r_{100} \approx 1.35\,r_{200}$. Here $\Delta_{v}$ is 
the ratio of the mean cluster mass density within $r_{\Delta_{v}}$ to the 
critical cosmic density. 

\section{Observations, Data Reduction \& Analysis}
\subsection{Observations}

\begin{figure}
\centerline{\FigureFile(60mm,150mm){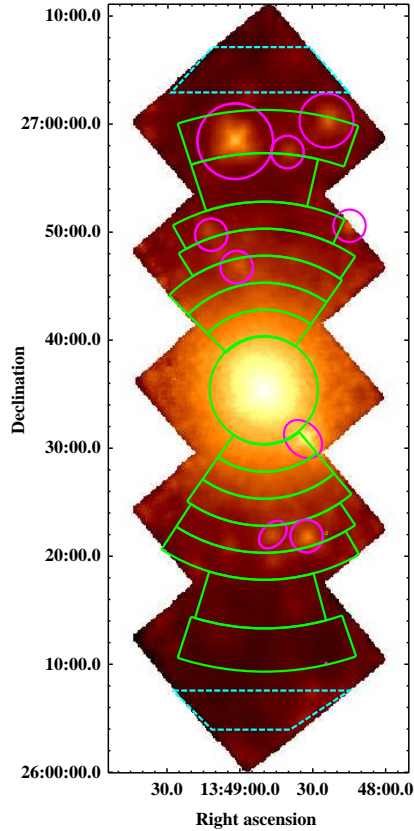}}
\caption{Mosaic of five pointings (0.5-2 keV), with North at top and East at
left. The image has been adaptively smoothed.  Cluster spectra are taken
from the solid green circle and annular sectors. Background regions are the
dashed blue trapezoids.  Magenta circles and ellipses surround excluded
point sources.}
\label{fig1}
\end{figure}

Abell 1795 was observed on 2005 Dec 10, during the Suzaku
performance verification phase,  with a series of 5 overlapping pointings,
illustrated in Figure~\ref{fig1}. The results reported here were obtained
with XIS~\citep{koy07} data from   version 2.0.6.13 of the Suzaku data
processing pipeline. Data obtained with 3x3 and 5x5 editing modes for each
pointing were merged and standard filters (requiring earth elevation
greater than 5 degrees, sunlit earth elevation angle greater than 20
degrees, cutoff rigidity greater than 6 GeV/c, and excluding data obtained
during and just after passages through the South Atlantic Anomaly) were
applied.  Events in detector regions nominally illuminated by the XIS
calibration sources were also excluded.  The net good exposure times after
filtering were 10.2 ks, 23.9 ks, 24.6 ks, 27.3 ks and 38.3 ks for the
central, near-north, near-south, far-north and far-south pointings,
respectively. Thus the total good exposure time was 124.3 ks.

Smoothed images of each field were inspected by eye and  regions  centered
on nine obvious point sources were excluded, as indicated in
Figure~\ref{fig1}. 

\subsection{Image Analysis}
\label{sexpmap}

To obtain a properly normalized cluster surface brightness profile from our
multiple pointings, we used the following procedure to generate  exposure
maps.  First, simulated monochromatic photon lists representing a uniform,
20\arcmin\ radius extended source were constructed, using 10$^7$ photons at a
number of energies for each detector and pointing.  The energy bins were
selected to represent roughly constant regions of the instrumental effective
area curve.  Each simulated photon list was used as input to the XRT
ray-tracing simulator ({\tt xissim}; \cite{ish07}), which includes the effects
of the optical blocking filter (OBF) contamination, producing an image for
each detector, pointing, and energy combination.  To increase signal-to-noise,
these flat field images were smoothed by a two dimensional Gaussian kernel
with 72\arcsec\ FWHM.  The calibration source regions and regions of low
effective area near the chip edges were clipped.  
Spectral weights were determined from a spectral model fit to the cosmic
X-ray background (described in Section~\ref{sec:cbg} below).  These
spectral weights were chosen under the null hypothesis that surface
brightness fluctuations in the cluster outskirts are due to cosmic
background variations.  When applied to cluster emission, they will result
in an overestimate of the cluster surface brightness, however we estimate
this to be less than 10\% (5\%) in the 0.5--2 keV (2--8 keV) range for the
cluster temperatures and abundances measured in this work.  
Errors in surface brightness between cluster regions of different
temperatures (ranging from 2--6 keV) will be less than 2\% due to the
spectral weighting.
For each chip and pointing, the flat field frames were combined using the
spectral weighting to produce an exposure map.  Finally, the individual
exposure maps were mosaicked onto a common sky coordinate (WCS) reference
frame and summed. 

Images from each pointing toward the  cluster were extracted in the 0.5--2 keV
and 2--8 keV bands, and these were mosaicked in sky coordinates.
Cutoff-rigidity-weighted particle background rates were computed for each
spectral band, sensor and pointing from the archive of `night-earth' data
(i.e., data obtained during orbit night while Suzaku  pointed at the
earth; \cite{taw08}).  A (spatially uniform) particle
background component was then subtracted from each cluster image, and each
resulting image was then  divided by the appropriate  exposure map.  The
exposure-corrected, 0.5--2 keV  mosaic image is shown in Figure \ref{fig1}. For purposes of this Figure, the background- and exposure-corrected image 
has been smoothed with an adaptive, circular kernel.  The kernel radius is 
chosen to include at least 25 counts, up to a maximum radius of 10 binned 
pixels ($ 84 \arcsec$).

\subsection{Spectral Analysis and X-ray Background}
\label{sec:spec_bg}

\subsubsection{Analysis and Cluster Modeling}
\label{sec:model}

Pulse-height spectra were accumulated for each in a series of annular
sectors 2.5--4\arcmin\ in width, depending on radius. The spectral
extraction regions are illustrated in Figure~\ref{fig1}.  
Two ancillary response functions (ARF files) were constructed for each
region and XIS detector using the Monte Carlo
FTOOL\footnote{http://heasarc.gsfc.nasa.gov/ftools} {\tt
xissimarfgen}~v2008-04-05 \citep{ish07}.  The first ARF was produced from a
20\arcmin\ radius uniform surface brightness source; this response file was
used to fit uniform background emission in the spectral analysis that
follows.  The second ARF was produced from the cluster surface brightness
distribution results described in Section \ref{sec:sb}; this response was
used to fit cluster emission in each region.  These files properly account
for energy-dependent vignetting and scattering properties of the 
Suzaku XRT, as well as for the time-dependent contamination layer on
each of the XIS optical blocking filters.  Corresponding detector
redistribution functions (RMF files) were constructed using the FTOOL {\tt
xisrmfgen}~v2007-05-14.  Cutoff-rigidity-weighted particle-induced
background spectra were determined from night-earth data for each sensor
and pointing using the FTOOL {\tt xisnxbgen}~v2008-03-08.  These background
spectra were subtracted from each source spectrum before fitting.

{\tt XSPEC} v12.5.0~\citep{ar96}\ was used to fit data from all four XIS
sensors simultaneously for all spectral fits reported here.  Spectral
fitting was confined to the 0.55--8 keV band where the response calibration
is best.
Calibration observations directly constrain the time-dependent, on-axis
absorption by contamination of the XIS optical blocking filter in this
spectral band~\citep{koy07}. 

\begin{figure*}
\centerline{\FigureFile(.8\linewidth,.8\linewidth){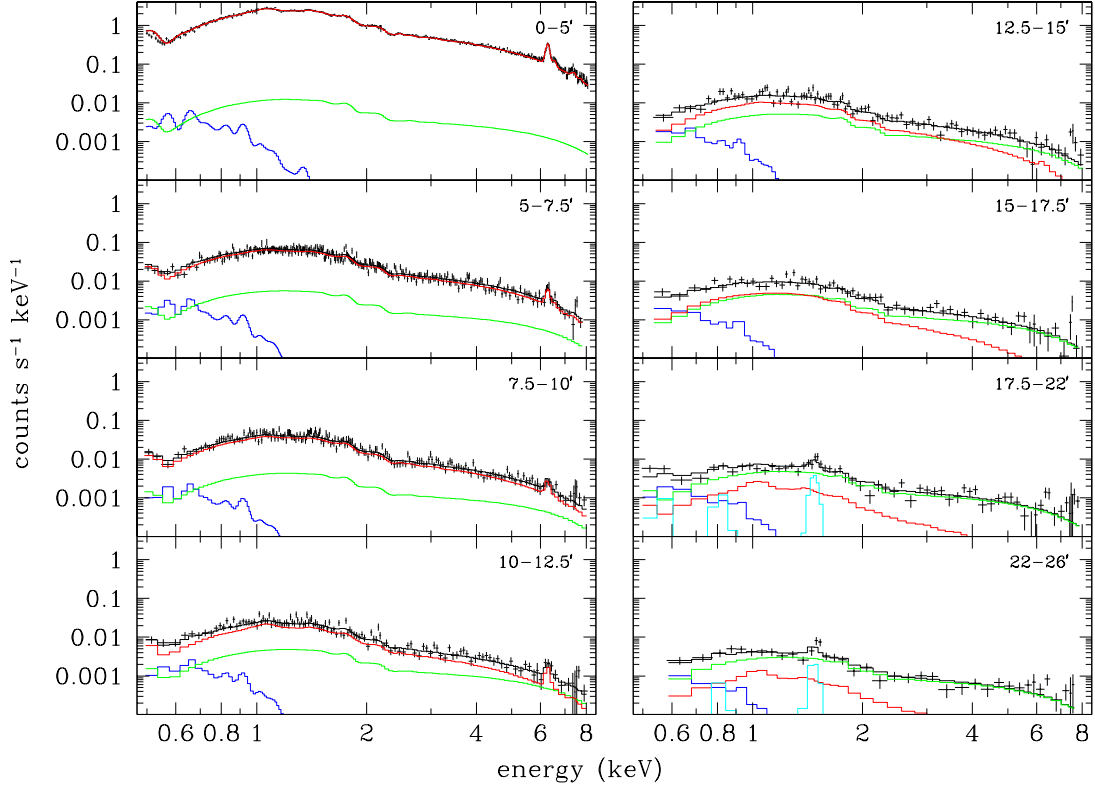}}
\caption
{Spectra of the central 5\arcmin\ radius region and annuli toward the
north from Abell 1795.  The locations of the regions are shown in
Figure \ref{fig1}.  For each region, the black points represent the
combined count rate from the three FI devices (XIS0,2,3), while the solid
lines show contributions from individual model components: the Galactic
background (blue); the extragalactic background (green); the cluster (red);
and solar wind charge exchange (cyan).  The total model is given by the
black line.}
\label{fig:northspec}
\end{figure*}

The emission  from each cluster region was modelled by simultaneously
fitting the pulse-height spectra from the cluster region and two background
regions.  The sum of a thin thermal plasma model (mekal in {\tt XSPEC})
plus a cosmic background model (described in Section~\ref{sec:cbg} below)
was fit to each  cluster spectrum, while only the background model was fit
to the background regions.  In all fits the cluster heavy element
abundance,  relative to cosmic values~\citep{ag89}, was allowed to vary.
The best-fit redshift determined from the central $5\arcmin$ region 
($z=0.0626 \pm 0.0008$) is consistent with the optical cluster redshift
($z=0.0625 \pm 0.0003$) reported by \citet{smi04}. The redshift was fixed at
the optical value for other spectral fits. 
The combined XIS0,2,3 (FI) spectra and best-fit model components for the
central 5\arcmin\ and the northern extension are shown in Figure
\ref{fig:northspec}.  

\subsubsection{Cosmic Background  Model}
\label{sec:cbg}

Spectra of the celestial X-ray background outside the cluster were obtained
from two trapezoidal background regions  at the northern and southern
extremes of the observed field, 2.0 to 2.3 Mpc from the cluster center; see
Figure~\ref{fig1}.  The background spectral model for each fit included the
sum of absorbed power-law and soft thermal components, to account for
extragalactic and Galactic backgrounds, respectively.  When modelling the
spectra of the faintest cluster regions (those with $r > 17.5\arcmin$) we
also included a suite of emission lines to account for possible
contributions from geocoronal solar wind charge exchange (SWCX).  We find
that the contribution of SWCX emission is uncertain and, though small, may
be significant in the interpretation of the emission detected in these
regions.  In this section we  describe the background model used for the
brighter ($r < 17.5\arcmin$) cluster regions, for which SWCX is neglected.
We  discuss our treatment of SWCX emission in the cluster outskirts in the
following section.

\begin{table*}
\begin{center}
\caption{Background Model Parameters$^a$\label{tab:fluxes}}
\begin{tabular}{lccc}
\hline
Component                & North                            & South                            & Joint \\
\hline
\multicolumn{4}{l}{without SWCX} \\
Power Law $I_{1\ keV}^b$ & 11.5 $^{+0.6}_{-0.7}$            & 10.3 $^{+0.6}_{-0.6}$            & 10.9 $^{+0.4}_{-0.4}$ \\
Thermal $kT^c$           & 0.15 $^{+0.03}_{-0.04}$          & 0.19 $^{+0.04}_{-0.02}$          & 0.18 $^{+0.02}_{-0.03}$ \\
Power Law $B^d$          & $\phantom{0}$7.7 $^{+0.4}_{-0.4}$  & $\phantom{0}$7.0 $^{+0.4}_{-0.4}$  & $\phantom{0}$7.3 $^{+0.3}_{-0.3}$ \\
Thermal $B^d$            & $\phantom{0}$2.9 $^{+0.7}_{-0.7}$  & $\phantom{0}$3.0 $^{+0.7}_{-0.6}$  & $\phantom{0}$3.0 $^{+0.5}_{-0.5}$ \\
Total $B^d$              & 10.6 $^{+0.7}_{-0.7}$            & 10.0 $^{+0.7}_{-0.7}$            & 10.3 $^{+0.5}_{-0.5}$ \\
\hline
\multicolumn{4}{l}{with SWCX} \\
Power Law $I_{1\ keV}^b$ & $11.3 ^{+0.6}_{-0.7}$            & 10.1 $^{+0.7}_{-0.7}$            & ... \\
Thermal $kT^c$           & $0.18 ^e$                        & $0.18 ^e$                        & ... \\
Power Law $B^d$          & $\phantom{0}$7.6 $^{+0.4}_{-0.5}$  & $\phantom{0}$6.8 $^{+0.5}_{-0.5}$  & ... \\
Thermal $B^{d,f}$        & $\phantom{0}$2.8 $^{+1.3}_{-2.5}$  & $\phantom{0}$2.8 $^{+1.1}_{-2.2}$  & ... \\
SWCX $B^{d,f}$           & $\phantom{0}$1.5 $^{+2.9}_{-1.0}$  & $\phantom{0}$1.3 $^{+2.3}_{-0.7}$  & ... \\
Total $B^{d}$            & 11.9 $^{+0.7}_{-0.8}$            & 11.2 $^{+0.7}_{-0.7}$            & ... \\
\hline
\multicolumn{4}{l}{\parbox{110mm}{\footnotesize
Notes: \\
\footnotemark[$a$] uncertainties are 90\% confidence for 1 interesting parameter\\ 
\footnotemark[$b$] intensity, ph s$^{-1}$ cm$^{-2}$ keV$^{-1}$ sr$^{-1}$ at 1 keV; photon index fixed at $\Gamma = 1.4$ \\
\footnotemark[$c$] keV; abundance fixed at solar \\
\footnotemark[$d$] surface brightness, $10^{-12}$ erg s$^{-1}$ cm$^{-2}$ deg$^{-2}$, 0.5--2 keV \\
\footnotemark[$e$] $kT$ was fixed at 0.18 keV for the model including SWCX \\
\footnotemark[$f$] the best-fit thermal and SWCX fluxes are anti-correlated
}}
\end{tabular}
\end{center}
\end{table*}

Best-fit spectral model parameters for the background regions, neglecting
SWCX emission,  are shown in the upper half of Table~\ref{tab:fluxes}.
Consistent results (within $\pm 6$\% in normalization)  were obtained for
each model component in  the two background regions. Since our spectral
analysis regions are quite small, expected Poisson fluctuations in the
extragalactic background are actually  larger than $\pm 6$\%, so this level
of agreement must be regarded as fortuitous.  To allow for these
fluctuations, the normalization of the corresponding spectral component was
allowed to vary separately when fitting each region of the cluster.  This
point is discussed in detail in Section~\ref{secbgva} below.

The 2--10 keV surface brightness of the power-law component, determined
from the joint fit to both the north and south background regions, is
$B_{2-10} = 21.7 \pm 1.5 \times 10^{-12}$ erg s$^{-1}$ cm$^{-2}$ deg$^{-2}$, 
in excellent agreement with the value reported by \citet{del04} from
XMM-Newton observations ($22.4 \pm 1.6 \times 10^{-12}$ erg
cm$^{-2}$ s$^{-1}$ deg$^{-2}$).  As a check on the estimated 0.5--2 keV
background, we determined the ROSAT All-Sky Survey (RASS) surface
brightness  in an annulus centered on Abell 1795 with inner and outer radii
of 0.5 and 1 degree, respectively.  We used the X-ray Background
Tool\footnote{{\tt http://heasarc.gsfc.nasa.gov/cgi-bin/Tools/xraybg/xraybg.pl}}~\citep{sno97}
provided by NASA's Goddard Space Flight Center to extract the ROSAT
spectrum for this region. We then fit the recommended X-ray background
model \citep{kun00}, which includes two soft thermal components and a 
power law, to the ROSAT spectrum in the 0.1--2 keV band. This yields a
surface brightness in the 0.5--2 keV band of $B_{0.5-2} = 11.1^{+2}_{-3}
\times 10^{-12}$ erg s$^{-1}$ cm$^{-2}$ deg$^{-2}$, in excellent agreement
with our (more precise) Suzaku estimates listed in
Table~\ref{tab:fluxes}. 

Thus, even ignoring possible contributions of SWCX emission,  our 
Suzaku background estimates are in excellent agreement with
previous measurements of both the Galactic and extragalactic components.
In our chosen spectral bands we can determine background (and foreground)
flux in the vicinity of Abell 1795 with a  relative  statistical
uncertainty  of just over 5\%. As we discuss in section~\ref{secbgva}
below, the accuracy  of our background subtraction is in fact  limited by
cosmic variations in the density of unresolved background sources rather
than by statistical or instrumental systematic errors.

\subsubsection{Solar Wind Charge Exchange Model}
\label{swcx}

Heavy ions in the solar wind undergo charge exchange with neutral atoms in
the Earth's geocorona and in the solar magnetosphere, resulting in a
field-filling emission line spectrum which contaminates the more distant
cluster emission.  Lines typically produced by solar wind charge exchange
(SWCX) include those of highly-ionized C, O, Ne, and Mg (e.g.,
\cite{Snowdenetal2004,Fujimotoetal2007}).  Variations in the solar wind
can produce variations in the geocoronal SWCX flux on timescales of
seconds, while heliospheric SWCX contributes a stable or slowly-varying
X-ray background (e.g., \cite{Cravens2000}).

\begin{figure}
\centerline{\FigureFile(.9\linewidth,.9\linewidth){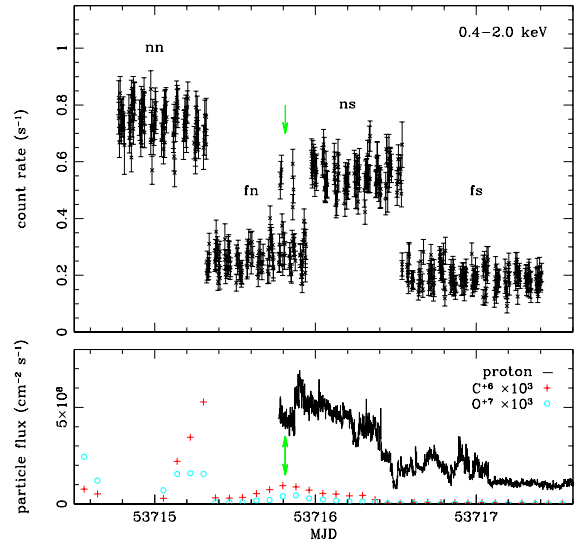}}
\caption
{({\em top\/}) Combined, background-subtracted light curve for all
detectors in the 0.4--2 keV range.  The four outer pointings are shown (nn:
near north; fn: far north; ns: near south; fs: far south).  ({\em
bottom\/}) Solar wind ion flux from the ACE satellite,
corrected for travel time to the Earth.  Count rate variation and
significant enhancements in the solar proton and ion flux indicate possible
SWCX contamination.  The green arrows indicate a strong count rate
flare.}
\label{fig:ace}
\end{figure}

We have checked for SWCX variability by constructing light curves for each
pointing, excluding the center pointing where the cluster emission is
expected to dominate any SWCX emission.  Light curves were extracted 
for all four XIS detectors in the 0.4--2 keV range from the full detector
field, excluding point sources.  Particle background light curves were
constructed from the `night-earth' data archive \citep{taw08} using the
appropriate cut-off rigidity time series, and these were subtracted from the
individual detector light curves, which were then combined for each field.
The resultant complete light curve is shown in the top panel of Figure
\ref{fig:ace}.  Any residual variations in count rate are a result of
variations in the emission from the Galactic foreground, the unresolved
extragalactic background, the cluster, and/or SWCX.  The extraction regions
for corresponding pointings are identical in sky coordinates, so any
differences between far north and far south (``fn'' and ``fs'') or between
near north and near south (``nn'' and ``ns'') must be due to variations
(temporal or spatial) in the surface brightness of these emission
components.

There is clear variability in the light curve.  Most notably, the far north
count rate shows a strong spike (marked by the green arrow).  Count rates
in the other three pointings are also inconsistent with the null hypothesis
of constancy (p-values $\lesssim$ 0.1), although they lack strong
variation.  Analysis of public data from the Advanced Composition
Explorer (ACE) satellite \citep{ACE} reveals an enhanced flux of
protons and heavy ions during much of the observation (see Figure
\ref{fig:ace}, bottom panel).  Although there are gaps in the ACE
data, the proton, C$^{+6}$, and O$^{+7}$ flux during the far north and near
north pointings (near MJD 53716) are similar to levels observed during
times of SWCX contamination in a Suzaku observation of the North
Ecliptic Pole \citep{Fujimotoetal2007}.  The far north count rate spike
occurs near a peak in both the proton and heavy ion flux.  There is an
additional spike in the heavy ion flux near the end of the near north
observation.  The solar particle flux near the end of the observations
(after MJD 53716.5) is consistent with that observed during quiescence in
other Suzaku observations \citep{Fujimotoetal2007,Milleretal2008}.
We conclude that our observations are likely contaminated by variable SWCX
emission.

We estimate the level and variability of geocoronal SWCX contamination by
comparing  the emission from the (presumed cluster-free) background
regions.  As in Section~\ref{sec:cbg} above, the non-SWCX background is
modeled as the sum of a soft thermal (Galactic) background component and a
power-law (extragalactic) component.  When fitting the spectra of the
relatively small background regions alone, we found strong degeneracies
between the SWCX and thermal model parameters, so we fixed the temperature
of the thermal model at $kT=0.18$ keV, the value determined from the
joint SWCX-free model.  The photon index of the power-law
component was fixed at $\Gamma = 1.4$.  The normalizations of both
background components were allowed to vary independently in each region
(see Section~\ref{secbgva}).  

To model the SWCX in the background regions, we include a suite of
unresolved Gaussian emission lines typically observed in SWCX emission
between 0.5--2 keV \citep{Snowdenetal2004,Fujimotoetal2007}, listed in
Table \ref{tab:swcxlus}.  The line energies are fixed to their theoretical
values, using the energy of the dominant forbidden transition for
helium-like lines of O~\textsc{vii}, Ne~\textsc{ix}, and Mg~\textsc{xi}
\citep{Koutroumpaetal2007}.  The line fluxes are allowed to vary
independently.  

\begin{table}
\begin{center}
\caption{Modeled SWCX Emission Lines \label{tab:swcxlus}}
\begin{tabular}{lcll}
\hline
transition & energy & \multicolumn{2}{c}{photon flux} \\
           & (keV)  & \multicolumn{2}{c}{(ph s$^{-1}$ cm$^{-2}$ sr$^{-1}$)} \\
           &        & North & South \\
\hline
O~\textsc{vii}  triplet     & 0.574 & 3.03 $^{+4.07}_{-1.56}$ & $<$ 6.66 \\
O~\textsc{viii} Ly$\alpha$  & 0.654 & $<$ 2.25                & $<$ 1.77 \\
O~\textsc{viii} Ly$\beta$   & 0.774 & $<$ 0.71                & $<$ 0.61 \\
O~\textsc{viii} Ly$\gamma$  & 0.817 & $<$ 0.63                & 0.49 $^{+0.50}_{-0.43}$ \\
Ne~\textsc{ix}  triplet     & 0.923 & $<$ 0.47                & 0.46 $^{+0.38}_{-0.35}$ \\
Ne~\textsc{x}   Ly$\alpha$  & 1.022 & $<$ 0.31                & 0.41 $^{+0.28}_{-0.27}$ \\
Mg~\textsc{xi}  triplet     & 1.345 & $<$ 0.11                & $<$ 0.23 \\
Mg~\textsc{xii} Ly$\alpha$  & 1.472 & 0.43 $^{+0.23}_{-0.22}$ & $<$ 0.16 \\
Mg~\textsc{xii} Ly$\beta$   & 1.745 & $<$ 0.08                & $<$ 0.11 \\
\hline
\end{tabular}
\end{center}
\end{table}

We use Markov chain Monte Carlo (MCMC) techniques to constrain the
relatively large number of model parameters and their uncertainties. 
The MCMC tools in {\tt XSPEC} v12.5.0 \citep{ar96} were used to produce 8
chains of 7000 steps, of which the first 2000 were excluded as ``burn-in''.
The input covariance matrix was scaled to get sufficient mixing (i.e., so
that the fraction of repeated locations in parameter space was $\sim$ 0.8).
These 40,000 chain steps were combined to produce probability distributions
for each of the 22 free parameters, and from these the best fit value and
90\% confidence intervals were estimated. 

\begin{figure*}
\begin{center}
\begin{minipage}[l]{.45\linewidth}
\centerline{\FigureFile(\linewidth,\linewidth){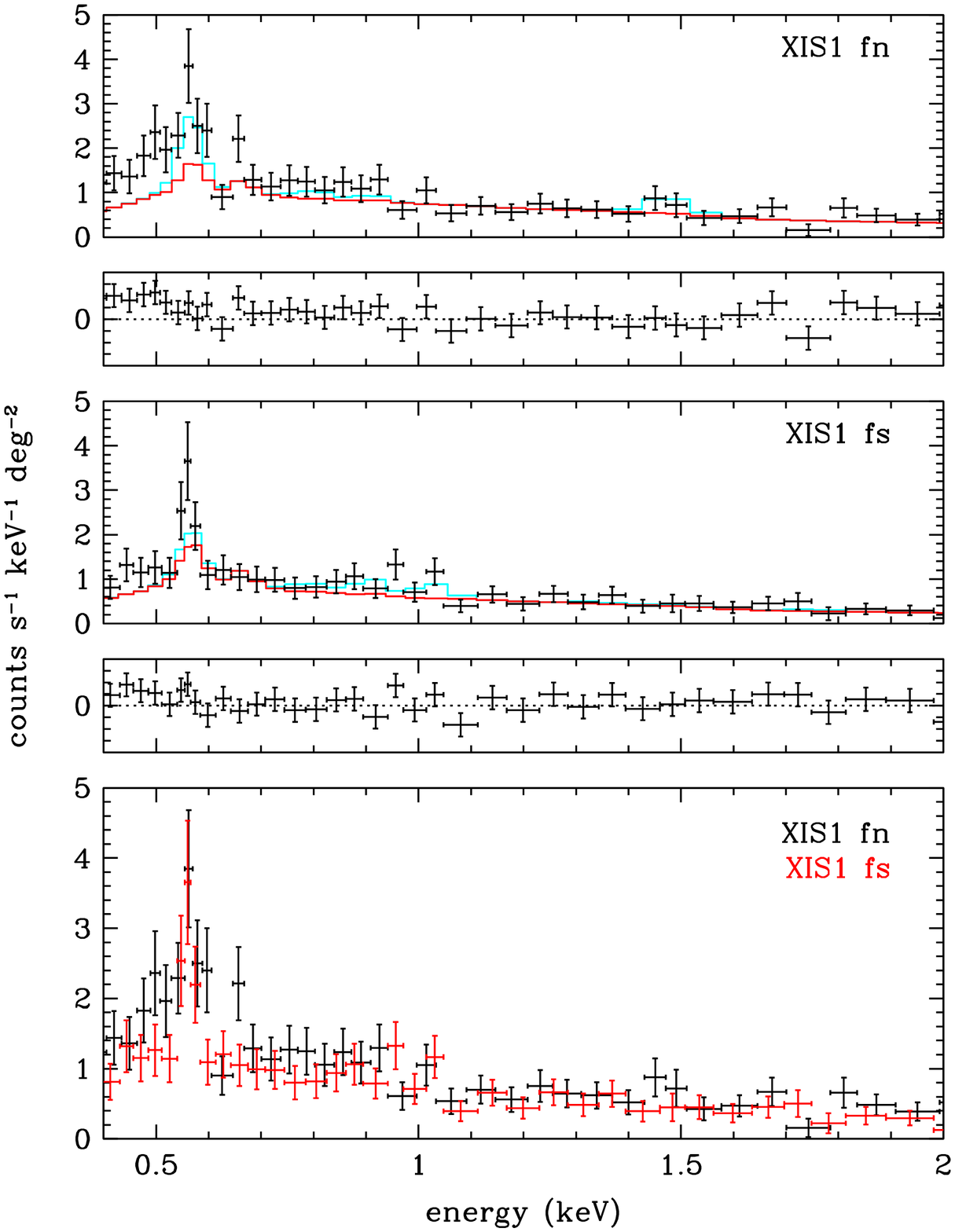}}
\end{minipage}
\begin{minipage}[r]{.45\linewidth}
\centerline{\FigureFile(\linewidth,\linewidth){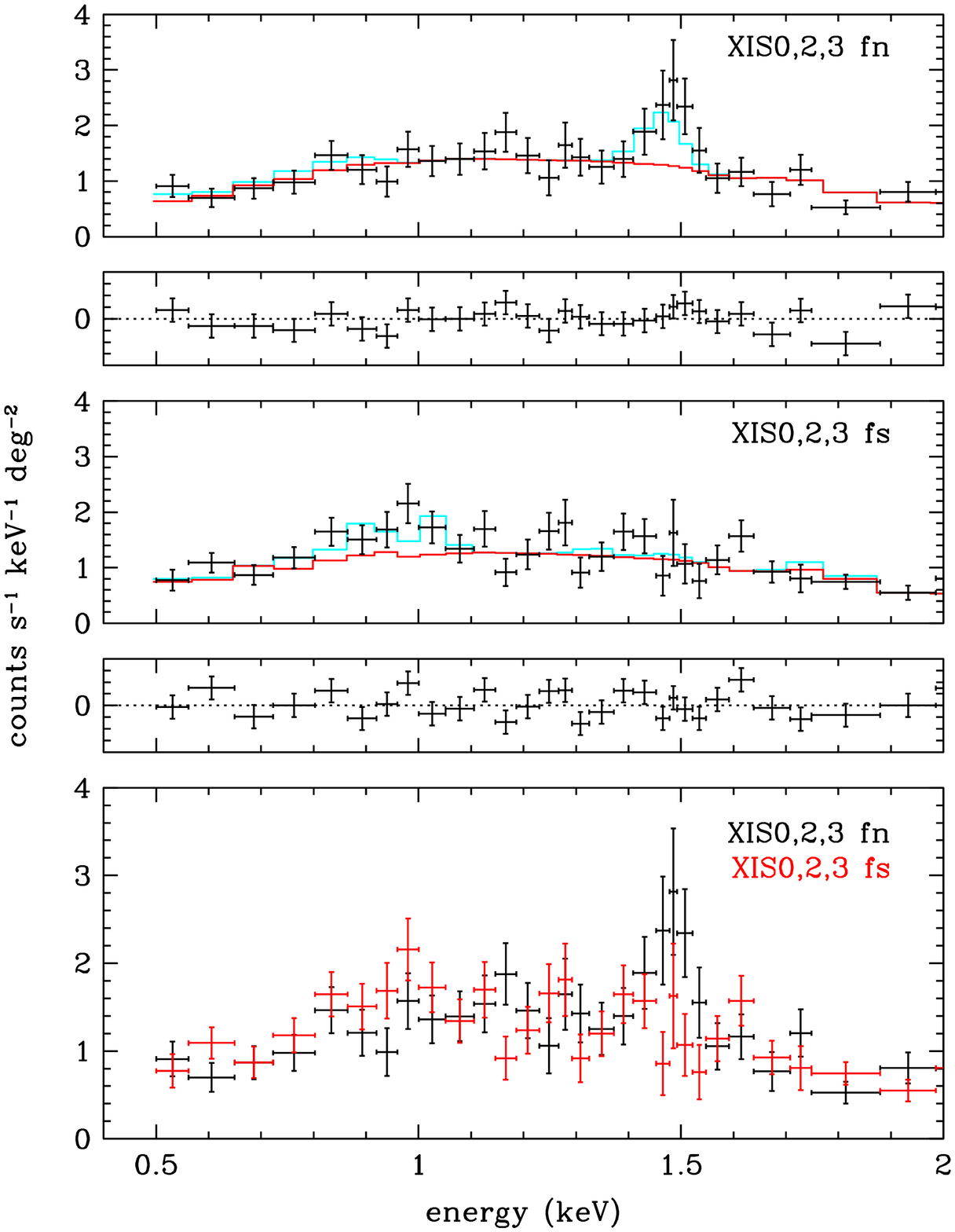}}
\end{minipage}
\end{center}
\caption
{Spectra of the two background regions from the BI detector XIS1 (left) and
the combined FI detectors XIS0,2,3 (right). The labels ``fn'' and ``fs''
refer to the far north and far south pointings, respectively.  In the top
panels, the red
lines indicate the best-fit diffuse Galactic + extragalactic background,
and the cyan lines indicate 90\% upper limits to SWCX emission modeled by
Gaussian lines (see Table \ref{tab:swcxlus}).  The residuals
are calculated from the combined background + SWCX model.  The model was
fit simultaneously to spectra from the four detectors; the FI spectra were
only combined here to improve the displayed S/N.  The bottom panels
directly compare the counts spectra from the far north (black) and far south
(red) background regions.  Differences can be seen near the O~\textsc{vii} 
(0.56 keV), Ne~\textsc{ix} and Ne~\textsc{x} (1 keV), and Mg~\textsc{xii}
(1.5 keV) lines.}
\label{fig:swcxspec}
\end{figure*}

Both the far north and far south background regions show an excess of line
emission above the Galactic and extragalactic components.  The spectra in
Figure \ref{fig:swcxspec} suggest enhancement of the
O~\textsc{vii} triplet near 0.56 keV (evident in XIS1)
and a $\sim$ 3-$\sigma$ feature near 1.5 keV (evident in XIS0,2,3) in the
far north spectrum.  The latter is consistent with Mg~\textsc{xii}
Ly$\alpha$.  The far south spectra show much weaker upper limits to the
enhancement near 0.56 keV (in XIS1) and 1.5 keV (in XIS0,2,3), but they
contain excess emission in the 0.9--1.0 keV region (evident in all
detectors) that is consistent with Ne~\textsc{ix} and Ne~\textsc{x}.  The
line strengths and upper limits are shown in Table~\ref{tab:swcxlus}.  The
Mg~\textsc{xii} line is similar in energy to a strong instrumental Al
K$\alpha$ line, however the Al K$\alpha$ line flux would have to be
enhanced by more than a factor of 3 during the full far north exposure to
produce such strong residual emission.  Background variations of this sort
have not been reported in previous Suzaku observations, nor in the
night-earth background analysis of \citet{taw08}.  The presence of a
Mg~\textsc{xii} SWCX line without additional SWCX lines such as
O~\textsc{viii}, Ne~\textsc{ix} and Ne~\textsc{x} seems puzzling, although
since detection of Mg~\textsc{xii} with XMM-Newton is complicated by a much
stronger and more variable instrumental Al K$\alpha$ line, there are few
data from other sources.  Mg~\textsc{xii} Ly$\gamma$ has likely been
observed in another XMM-Newton SWCX event \citep{CarterSembay2008}.
\citet{Snowdenetal2004} claim a Mg~\textsc{xi} SWCX detection at 1.34 keV.
The upper limits on the photon fluxes for other expected lines are low
compared to other SWCX detections, which suggests that our observations are
less affected by geocoronal SWCX.  

The contributions  of SWCX and other  model components to the total surface
brightness in the background regions are listed in the lower half of
Table~\ref{tab:fluxes}.  The Table shows that SWCX emission is detected at
better than 90\% confidence during our observations of both far north and
far south fields, and that the SWCX emission spectrum  differs significantly
between the two observations. However,  in each field SWCX  contributes less
than 15\% of the total, soft-band cosmic background brightness, and the
difference between SWCX contributions to the two fields is less than 5\% of
the total cosmic background brightness. 

Our spectral analysis of the outermost regions of the cluster 
($r > 17.5\arcmin$) follows that used for the interior of the cluster,
except that we include the SWCX model described above in the background
model (see Figure \ref{fig:northspec}).  In particular, the soft thermal
(Galactic) background component was fit with a single free temperature and
normalization for all regions (17.5--$26\arcmin$ north and south, and the
north and south background regions).  The power-law (extragalactic)
background component was fit with a fixed photon index $\Gamma = 1.4$ and
normalizations that were allowed to vary independently in each region (see
Section~\ref{secbgva}).  A mekal component, with temperature, abundance,
and normalization free to vary, was used to model cluster emission.

\section{Systematic Errors}

We wish to measure cluster emission at very low surface brightness levels
and it is important to understand the influence of various instrumental and
other systmatic effects that may limit our ability to do so.  In this
section we consider a number of potential systematic errors in more detail,
including i) the impact of our choice of passband on temperature
measurement accuracy; ii) the accuracy of our knowledge of spatial response
variations  over the XIS field of view; iii) the impact of scattered light,
both from the bright cluster core and from point sources in the field of
view, on the accuracy of our spectral modelling; and iv) the effects of
counting statistics on the flux of unresolved sources. 

\subsection{Spectral passband, Galactic absorption and cluster temperature}

A drawback of our decision to ignore data at energies below 0.55 keV in our
spectral fits is that it is difficult to constrain the Galactic absorption
directly from the X-ray data unless the signal-to-noise ratio is quite
high.  On the other hand, we have found that the best-fit model temperature
and Galactic column do exhibit a mild (anti) correlation.  Thus, any
systematic errors in determination of the Galactic column (such as might
arise from spatially varying errors in the contamination correction) can
induce errors in the measured temperature.  We have therefore elected to
fix the model Galactic column density at the value obtained by fitting to
the high-signal-to-noise ratio spectrum of the integrated cluster emission
at $r < 5.0\arcmin$. Fitting results are shown in Table~\ref{tab2}.  The
best-fit value of $n_{H}$ is in excellent agreement with the nominal
Galactic column ($n_{H} = 1.2 \times 10^{20}$ cm$^{-2}$) estimated for this
field by NASA's HEARSARC tool {\tt nH}.\footnote{{\tt
http://heasarc/gsfc.nasa.gov/cgi-bin/Tools/w3nh}. The column densities
estimated from the Leiden/Argentine/Bonn \citep{kal05} and \citet{dl90} maps
agree within 1\%.}

\begin{table*}
\caption{Best-fit Model Parameters for $r < 5.0\arcmin$} \label{tab2}
\begin{center}
\begin{tabular}{ccccc}
\hline
Flux$^{a}$            & $kT^{b}$               & Abundance         & $n_H\,^{c}$   & $\chi^{2}$/DOF \\
\hline
$7.46^{+0.04}_{-0.03}$   & $5.26 \pm 0.06$        & $0.360 \pm 0.018$ & $1.2 \pm 0.3$ & $4709.6/4481$\\
\hline
\multicolumn{5}{l}{\parbox{110mm}{\footnotesize
Notes: \\
\footnotemark[$a$] $10^{-11}$ erg s$^{-1}$ cm$^{-2}$, 0.55--8 keV;
\footnotemark[$b$] keV;
\footnotemark[$c$] $10^{20}$ cm$^{-2}$
}}
\end{tabular}
\end{center}
\end{table*}

\subsection{Spatial response variations and reproducibility of spectral
modelling}

\begin{figure}
\centerline{\FigureFile(\linewidth,\linewidth){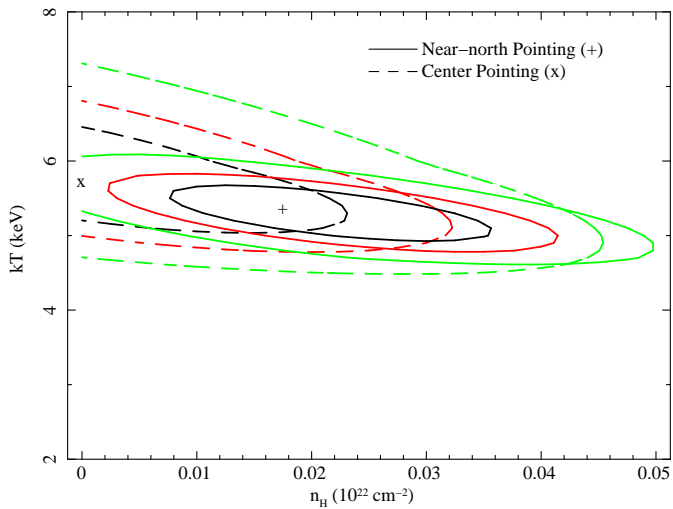}}
\caption{Contours of $\chi^{2}$ over the n$_{H}$-$kT$ plane derived from
two distinct observations of region of Abell 1795. The region in question
is an annular sector 5\arcmin--7.5\arcmin\ north of the cluster.  Contours at
68\% (black), 90\% (red) and 95\% (green) confidence from the `near-north'
(solid contours) and `center' (dashed contours) pointings are indicated.
The cross and x symbols indicate the best fit values for these two regions,
respectively. The good agreement of these measurements from opposite sides
of the XIS field of view illustrates the accuracy with which spatial
response variations have been calibrated.}
\label{fig_nhkT}
\end{figure}

As a check on the reliability of the Suzaku calibration, we
compared fitting results for regions of the cluster observed with multiple
pointings. Two such `overlap' regions provide  sufficient flux for a
meaningful comparison.  Each region is an annular sector with inner and
outer radii of $5.0\arcmin$ and $7.5\arcmin$, respectively.  The northern
overlap region was observed with central and `near-north' pointings; the
southern overlap region was observed with the central and `near-south'
pointings. Thus each overlap region was observed on opposite sides of the
XIS field of view. In order to gauge errors in the calibration of the soft
response, we allowed the  Galactic column to vary when fitting individual
spectra.  For each overlap region, the best-fit model parameters from the
two pointings are consistent with one another within 90\% confidence
statistical errors ($\delta kT = 0.3$ keV). The level of agreement, as well
as the magnitude of calibration errors, is illustrated in
Figure~\ref{fig_nhkT}, which shows contours of $\chi^{2}$ in the 
$n_{H}$-$kT$ plane for the northern overlap region.  Figure~\ref{fig_nhkT}
suggests residual systematic errors in the determination of the absorbing
column may be  of order  $\delta n_{H} \sim 1 \times 10^{20}$ cm$^{-2}$. As
noted above, in subsequent analysis we fix $n_{H} = 1.2 \times 10^{20}$
cm$^{-2}$ and include a systematic temperature error of $\delta kT = 0.15$
keV to account for this effect. 

\subsection{Scattered X-ray flux}

We estimated the effect of X-rays scattered by the XRT from  the bright,
cool core of Abell 1795 to the cluster outskirts with the aid of a number
of simulations. For this purpose we used the FTOOLS {\tt xissim} and {\tt
xissimarfgen}~\citep{ish07}.  For fields centered $20\arcmin$ from a bright
source (comparable to our far north and far south fields), actual stray
intensity levels are expected to be less than $\sim 2$ times the values
predicted by these tools \citep{ser07}.

To gauge the effect of scattered X-rays from the bright cluster core on the
surface brightness profiles at large radius, we simulated observations of a
point source located at the cluster center with a spectrum and total flux
observed  within $5\arcmin$ of the center (see Table~\ref{tab2}).  Since
the actual cluster emission is more diffuse than assumed in this model, our
simulation  overestimates the scattered flux from the (unresolved) bright
cool core of the cluster.  We simulated the full mosaic observation of five
pointings for each sensor, and normalized the simulated observations using
the exposure maps discussed in Section~\ref{sexpmap} above.  We then
extracted surface brightness profiles for this point source using the same
methods we used to extract the cluster profiles (as discussed in Section
\ref{sec:sb}). In particular, all
regions excluded from the cluster profiles because they contained cosmic
sources were also excluded in the simulations. For clarity, we did not add
a cosmic X-ray background signal to the simulated point source profile.

\begin{figure}
\centerline{\FigureFile(\linewidth,\linewidth){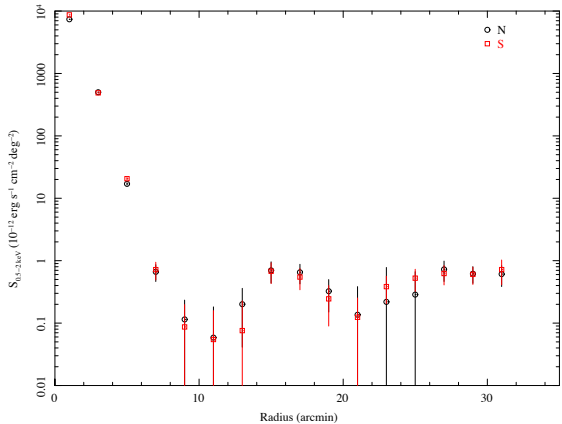}}
\caption{Simulated surface brightness (points) in the 0.5--2 keV band for a
point source located at the cluster center with flux  equal to that from
within $r < 5\arcmin$ of the cluster center. Surface brightness  northward
(black circles) and southward (red squares) from the source location are
shown.  No cosmic background has been added.}
\label{figsimps}
\end{figure}

The resulting point-source surface brightness profile  in the 
0.5--2 keV band is shown by the points in Figure~\ref{figsimps}.  The
Figure shows that the expected scattered flux  at $r > 10\arcmin$ exhibits
a distinct oscillation at relatively low amplitude.  The oscillation is
presumably due to the variations with field angle of the intensity of
so-called `secondary' (single bounce) and `backside' (three-bounce)
reflections~\citep{ser07}.

Comparison of Figure~\ref{figsimps} with the estimated celestial background
brightness listed in Table~\ref{tab:fluxes} shows that the surface brightness
due to scattered flux in this radial range is always less than 7\% of the
cosmic background surface brightness we observe at the largest radii ($r >
30\arcmin$).  The actual scattered intensity from the cluster will be below
that shown in Figure~\ref{figsimps} at these radii because convolution of
the point response function (PRF) with the extended cluster flux
distribution will reduce the amplitude of the oscillating wings.
\citet{ser07} note that, given
alignment tolerances of the XRT reflectors and pre-collimator blades, the
actual scattered flux at $r = 20\arcmin$ could in principle exceed the
ray-tracing simulations (which assume  perfect alignment) by as much as a
factor of 2. Allowing for this uncertainty,  we expect that scattered
cluster flux will contribute at most 10\% of the total observed brightness
at $r > 10\arcmin$. 

We confirmed the effect of scattered cluster flux on our spectral
analysis in the outermost regions of our field ($17.5\arcmin < r <
32\arcmin$)  by computing the effective area (as a function of energy) for
detection of photons there originating from the central $5\arcmin$ of the
cluster.  
Here we used the measured cluster surface brightness distribution as input
to the effective area calculation, rather than a simple point source.
Note that this central region produces more than
80\% of the  total cluster flux reported for A1795  in the Northern ROSAT
All-Sky Survey (NORAS) catalog~\citep{boe00}. Using these effective area
functions, and the  cluster spectral model described in Table~\ref{tab2}, we
employed {\tt XSPEC} to simulate the flux and spectrum of X-rays scattered
from the cluster center into each cluster (and background) region at $r >
17.5\arcmin$.  The resulting fluxes are listed in the first two columns of  Table~\ref{tab3}.  

\begin{table*}
  \caption{Estimated scattered flux} \label{tab3}
  \begin{center}
    \begin{tabular}{l c c c c c c}
      \hline
           &   \multicolumn{2}{c}{From Cluster Core} &  \multicolumn{2}{c}{From N. Point Source}   & \multicolumn{2}{c}{Total} \\
Region$^{a}$ &   $B_{sc,0.5-2}^{b}$ & $B_{sc,2-8}^{b}$ & $B_{sc,0.5-2}^{b}$ & $B_{sc,2-8}^{b}$ &   $B_{sc,0.5-2}^{b}$ & $B_{sc,2-8}^{b}$ \\
      \hline
      17.5--22   (N)  & 0.36    & 0.25  &   0.38  & 0.02 & 0.74 & 0.27 \\
      17.5--22   (S)  & 0.21    & 0.16  &   ...   & ...  & 0.21 & 0.16 \\
                      &         &       &         &      &       \\
      22--26     (N)  & 0.37    & 0.33  &   0.45  & 0.03 & 0.82 & 0.36   \\
      22--26     (S)  & 0.37    & 0.36  &   ...   & ...  & 0.37 & 0.36    \\
                      &         &       &         &      &       \\
      27.5--31.7 (N)  & 0.47    & 0.35  &   ...   & ...  & 0.47 & 0.35  \\
      27.5--31.7 (S)  & 0.45    & 0.34  &   ...   & ...  & 0.45 & 0.34    \\
      \hline
\multicolumn{7}{l}{\parbox{110mm}{\footnotesize
Notes: \\
\footnotemark[$a$] Minimum and maximum radii, arcmin, north (N) or south (S) of cluster center. \\
\footnotemark[$b$] $10^{-12}$ erg s$^{-1}$ cm$^{-2}$ deg$^{-2}$, 0.5--2 and 2--8 keV \\
}}
\end{tabular}
\end{center}
\end{table*}

We conclude from Table~\ref{tab3}  that the scattered intensity from the
cluster core  contributes  $\sim$ 3--5\%  of the total observed surface
brightness in the regions for which we perform spectroscopy. Both magnitude
and variation with radius are  in  agreement with the foregoing analysis of
the PRF.   We reiterate that uncertainties in collimator alignment render
these estimates uncertain by as much as a factor of 2.  The estimated
scattered fluxes in the northern and southern regions differ somewhat,
particularly in the radial range 17.5--22\arcmin.  The differences are
presumably due to the exclusion of point sources in the north (see
Figure~\ref{fig1}), which changes the effective radii of the annular bins
there, coupled with the rapidly varying scattered flux in this region (see
Figure~\ref{figsimps}).  In any event, the scattered flux in all regions
listed in Table~\ref{tab3} is comparable to or less than  the  statistical
errors in the estimated background (see Table~\ref{tab:fluxes}). We note
that, because of the oscillating PRF wings (see Figure \ref{figsimps}), the
simulations imply that
spatial non-uniformity of the scattered flux  will  cause us to
over-estimate the background at $r > 27.5\arcmin$, and thus to (slightly)
underestimate the cluster flux in the outer-most regions.  

Finally, we investigated the effect of scattered flux from  bright point
sources on our cluster spectrophotometry.  The brightest such source is in
the far-north field, with Suzaku position\footnote{This source is
coincident, within expected Suzaku position error of up to 20
arcsec, with 1RXS J134903.6+265845.} $\alpha = 13{\rm h} 49{\rm m} 03.0{\rm
s}, \delta = +26^{\circ} 58^{\prime} 46.1^{\prime \prime}$, and we estimate
its flux to be $4.0 \pm 1.9 \times 10^{-13}$ erg s$^{-1}$ cm$^{-2}$ (0.5--2
keV) within an aperture of $2.5\arcmin$ radius. We have excluded a 
3.5\arcmin\ diameter region around this source from the cluster analysis. We
estimated the flux contributed by this source to each of our spectroscopic
regions following methods described above for the cluster flux.  Results
are listed in Table~\ref{tab3}. The wings of the image of this  bright
point source contribute less than 5\% of the observed total flux to each of
the two outermost spectral analysis regions (extending from $17.5\arcmin$
to $26\arcmin$ from the cluster). Thus in these two regions, the bright
point source contribution is comparable to that from scattered cluster
flux.  In all other regions, the bright point source contribution is
smaller. We note that point source contributions should be  known more
accurately than those from the cluster core because the the ray-tracing
simulations are better calibrated at $r \sim 3\arcmin$--$10\arcmin$ than at $r >
20\arcmin$ \citep{ser07}.

\subsection{Cosmic background variations}
\label{secbgva}

Wide and deep X-ray surveys conducted over the past $\sim 15$ years have
revealed a great deal about the distribution in flux and space of the
sources that constitute the extragalactic X-ray background (\cite{bh05} and
references therein). We have used this information to estimate the expected
variation of the extragalactic background not resolved by Suzaku.
On scales larger than the Suzaku PRF ($\sim 2\arcmin$), the
correlation of source angular positions is weak~\citep{ya03}, and for
purposes of estimating the expected variance in our background, we neglect
it.  Formally the expected background surface brightness $B$ due to sources
with flux $S < S_{excl}$ is $B = \int_{0}^{S_{excl}} \frac{dN}{dS} S dS$,
where $\frac{dN}{dS}$ is the differential number of sources per unit solid
angle and flux.  The expected variance of the background brightness,
$\sigma_{B}^{2}$, when measured over a solid angle $\Omega$, is
$\sigma_{B}^{2} = (\int_{0}^{S_{excl}} \frac{dN}{dS} S^{2} dS)/\Omega$.  We
adopt a broken power-law model for $\frac{dN}{dS}$ using parameters derived
by~\citet{mo03}.  

It may be instructive to note that if the integral number counts
$N(S>S_{excl})$ varied simply as $N(S>S_{excl}) \propto S_{excl}^{- \alpha}$,
then the relative fluctuation in the the unresolved background within a
solid angle $\Omega$ would satisfy  $\sigma_{B}/B =
K(\alpha)/(N(S>S_{excl})\Omega)^{1/2}$, where the factor $K(\alpha)$ varies
between $\sim 1/2$ and $3/2$ for appropriate values  of $\alpha$.  This
relationship sets the flux limit $S_{excl}$  to which point sources must be
detected  for accurate background estimation in a field of  solid angle
$\Omega$. 

Our Suzaku data alone allow us to detect  point sources to a
limiting flux $S >  S_{excl} = 10\,S_{14}$, where $S_{14}$ is the flux in
units of $10^{-14}$ erg s$^{-1}$ cm$^{-2}$, in either the soft (0.5--2
keV) or hard (2--8 keV) bands.  For a significant fraction of our field, we
can also use archival XMM-Newton observations to constrain the number of
unresolved sources contributing to the Suzaku  background at fluxes
as low as
$S_{excl} = 1\, S_{14}$ in the soft band, and 
$S_{excl} = 1.3\, S_{14}$  in the hard band (see below).
Using our adopted source-count flux relation, we find the expected RMS
fluctuations in background due to unresolved sources are 
$\sigma_{B} = 3.9 (0.94)\times 10^{-12}\, \Omega_{0.01}^{-1/2}$ 
erg s$^{-1}$ cm$^{-2}$ deg$^{-2}$ for 
$S_{excl} = 10(1)\,S_{14}$ in the soft band.
The corresponding fluctuation amplitudes in the hard  band are
$\sigma_{B} = 4.5 (2.5)\times 10^{-12}\, \Omega_{0.01}^{-1/2}$ 
erg s$^{-1}$ cm$^{-2}$ deg$^{-2}$ for 
$S_{excl} = 10(1.3)\,S_{14}$.
Here $\Omega_{0.01}$ is the solid angle of the measurement region 
in units of $10^{-2}$ deg$^{2}$. 

\begin{figure}
\centerline{\FigureFile(60mm,150mm){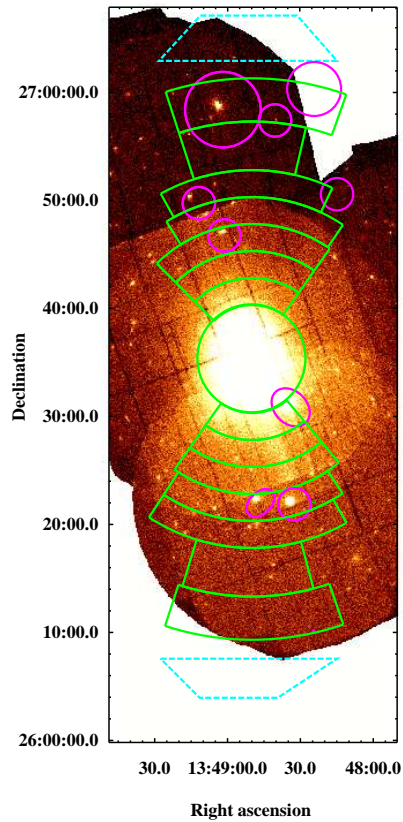}}
\caption{XMM-Newton image of Abell 1795 field with Suzaku spectral
analysis regions superposed in green.  Where available, the XMM-Newton
data constrain the background uncertainties due to point sources unresolved
by Suzaku.  As in Figure~\ref{fig1}, background regions are the dashed
blue trapezoids and magenta circles and ellipses show point sources
excluded from the Suzaku data.}
\label{figxmm}
\end{figure}

The field coverage of the available XMM-Newton observations is shown in
Figure~\ref{figxmm}.  We searched the Second XMM-Newton Serendipitous Sky
Survey (XMM2P) catalog available at  NASA's High Energy Astrophysics
Science Archive Research Center\footnote{{\tt http://heasarc.nasa.gov/}}
for sources in our Suzaku field at distance $10\arcmin < r < 26\arcmin$
from the cluster center.  We found that the effective (EPIC PN) exposure in
this region ranges from 10--50 ksec. Our adopted XMM-Newton flux
thresholds ($S_{excl} = 1\,S_{14}$ and $S_{excl} = 1.3\,S_{14}$ in the soft
and hard bands, respectively) exceed the nominal XMM-Newton $5\sigma$
detection thresholds \citep{wat01} for the minimum effective exposure
available in our field.  We summed the flux of sources brighter than these
thresholds in each of our spectral analysis regions. Since the XMM2P
catalog does not tabulate source flux in the 2--8 keV band, we estimated
the 2--8 keV flux by scaling the tabulated XMM-Newton 2--12 keV flux by a
factor appropriate for power-law photon number distribution with index 1.4
(that is, we assumed $S_{2-8} = S_{2-12} \times 0.67$.) The contributions
of sources resolved by XMM-Newton to the Suzaku background, as
well as the estimated fluctuations in the background due to fainter
sources, are listed in Table~\ref{bgtab}.

\begin{table*}
\caption{Estimated Background Contributions from Point Sources\label{bgtab}}
\begin{center}
\begin{tabular}{l c c c c c c c c c c c}
\hline
           &       &             & \multicolumn{4}{c}{0.5--2 keV}    &             & \multicolumn{4}{c}{2--8 keV} \\
Radius$^{a}$ & $\Omega_{0.01}^{b}$ &      & $S_{excl}^{c}$ & $B_{res}^{d}$ & $B_{unres}^{d}$ & $\sigma_{unres}^{d}$ &    & $S_{excl}^{e}$ & $B_{res}^{f}$ & $B_{unres}^{f}$ & $\sigma_{unres}^{f}$ \\
\hline
North:     &       & \phantom{0} &      &          &      &          & \phantom{0} &         &         &       & \\
5--7.5     & 0.60  & \phantom{0} & $10$ & 0$^{g}$  & 7.8  & 4.9      & \phantom{0} & $10$    & 0$^{g}$ & 16.4  & 5.8\\
7.5--10    & 0.85  & \phantom{0} & $10$ & 0$^{g}$  & 7.8  & 4.2      & \phantom{0} &  $10$   & 0$^{g}$ & 16.4  & 4.9 \\
10--12.5   & 0.93  & \phantom{0} & $1$  & 2.2      & 3.1  & 0.96     & \phantom{0} & $1.3$   & 2.7     & 12.0  & 2.6 \\
12.5--15   & 0.98  & \phantom{0} & $1$  & 3.9      & 3.1  & 0.94     & \phantom{0} & $1.3$   & 3.8     & 12.0  & 2.6 \\
15--17.5   & 0.91  & \phantom{0} & $1$  & 1.3      & 3.1  & 0.98     & \phantom{0} & $1.3$   & 0.      & 12.0  & 2.6 \\
17.5-- 26  & 1.58  & \phantom{0} & $1$  & 0        & 3.1  & 0.74     & \phantom{0} & $1.3$   & 7.4     & 12.0  & 0.8 \\
27.5--31.7 & 1.54  & \phantom{0} & $10$ & 0$^{g}$  & 7.8  & 3.1      & \phantom{0} & $10$    & 0$^{g}$ & 16.4  & 3.6 \\
\\
South:     &       & \phantom{0} &      &          &      &          & \phantom{0} &         &         &       & \\
5--7.5     & 0.57  & \phantom{0} & $10$ & 0$^{g}$  & 7.8  & 5.1      & \phantom{0} & $10$    & 0$^{g}$ & 16.4  & 6.0 \\
7.5--10    & 0.79  & \phantom{0} & $10$ & 0$^{g}$  & 7.8  & 4.3      & \phantom{0} & $10$    & 0$^{g}$ & 16.4  & 5.1 \\
10--12.5   & 0.93  & \phantom{0} & $1$  & 0        & 3.1  & 0.96     & \phantom{0} & $1.3$   & 0       & 12.0  & 2.6 \\
12.5--15   & 0.83  & \phantom{0} & $1$  & 1.3      & 3.1  & 1.0      & \phantom{0} & $1.3$   & 4.6     & 12.0  & 2.7 \\
15--17.5   & 1.22  & \phantom{0} & $1$  & 2.0      & 3.1  & 0.84     & \phantom{0} & $1.3$   & 3.8     & 12.0  & 2.3 \\
17.5--26   & 3.11  & \phantom{0} & $1$  & 0        & 3.1  & 0.54     & \phantom{0} & $1.3$   & 1.1     & 12.0  & 1.4 \\
27.5--31.7 & 1.25  & \phantom{0} & $10$ & 0$^{g}$  & 7.8  & 3.5      & \phantom{0} & $10$    & 0$^{g}$ & 16.4  & 2.2 \\
\hline
\multicolumn{12}{l}{\parbox{140mm}{\footnotesize
Notes: \\
\footnotemark[$a$] Radial range of region relative to cluster center, arcmin; 
\footnotemark[$b$] Region solid angle, $10^{-2}$ deg$^{2}$; 
\footnotemark[$c$] Limiting flux for resolved sources, in units $10^{-14}$ ergs s$^{-1}$ cm$^{-2}$, 0.5--2 keV; 
\footnotemark[$d$] $10^{-12}$  ergs s$^{-1}$ cm$^{-2}$ deg$^{-2}$, 0.5--2 keV; 
\footnotemark[$e$] Limiting flux for resolved sources, in units $10^{-14}$ ergs s$^{-1}$ cm$^{-2}$, 2--8 keV;  
\footnotemark[$f$] $10^{-12}$  ergs s$^{-1}$ cm$^{-2}$ deg$^{-2}$, 2--8 keV; 
\footnotemark[$g$] These regions lack useful XMM-Newton coverage.
}}
\end{tabular}
\end{center}
\end{table*}

In the outskirts of the cluster ($r > 10\arcmin$), the expected
fluctuations in the unresolved background are $\approx 10^{-12}$ erg
s$^{-1}$ cm$^{-2}$ deg$^{-2}$ (0.5--2 keV), or about 10\% of the total
(Galactic plus extragalactic) background.  In our two background regions at
$r > 27.5\arcmin$, which lack complete XMM-Newton coverage, the expected
fluctuations are significantly larger, ranging from 
3--$3.5 \times 10^{-12}$ erg s$^{-1}$ cm$^{-2}$ deg$^{-2}$. 
We note also that the expected extragalactic background brightness for
these regions is in excellent agreement with the measured values (see Table
\ref{tab:fluxes}). 

Finally, we stress that the expectation value of the extragalactic
background brightness varies from region to region because the
extragalactic source population (as well as our knowledge of it) varies
(see Table~\ref{bgtab}).  Thus, for example, it is  appropriate when
modelling the background to allow  the extragalactic component to vary in
brightness from region to region, and we do so in our fits.

\section{Results and Discussion}

\subsection{Surface Brightness}
\label{sec:sb}

\begin{figure*}
\centerline{\FigureFile(4.5in,3.375in){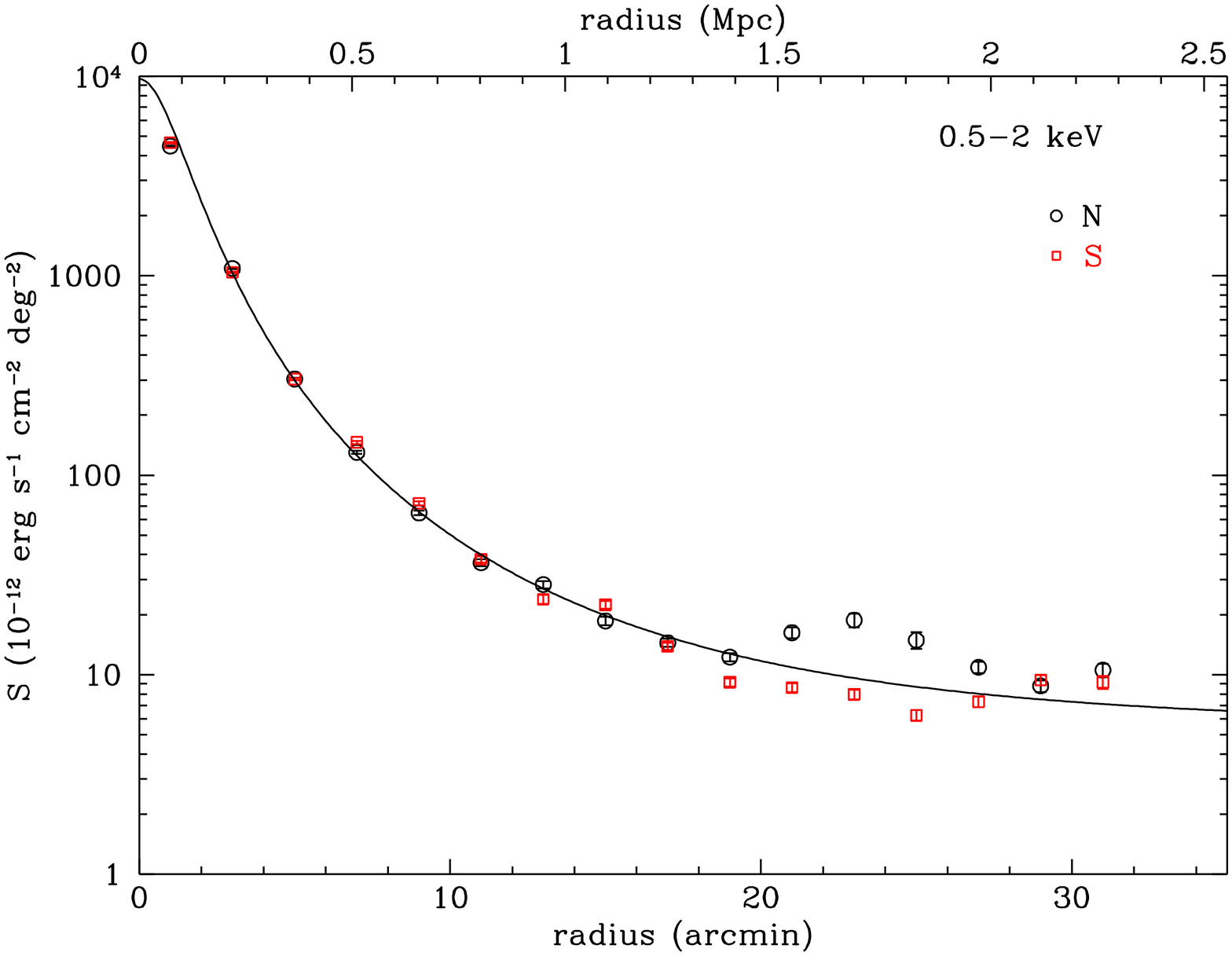}}
\centerline{\FigureFile(4.5in,3.375in){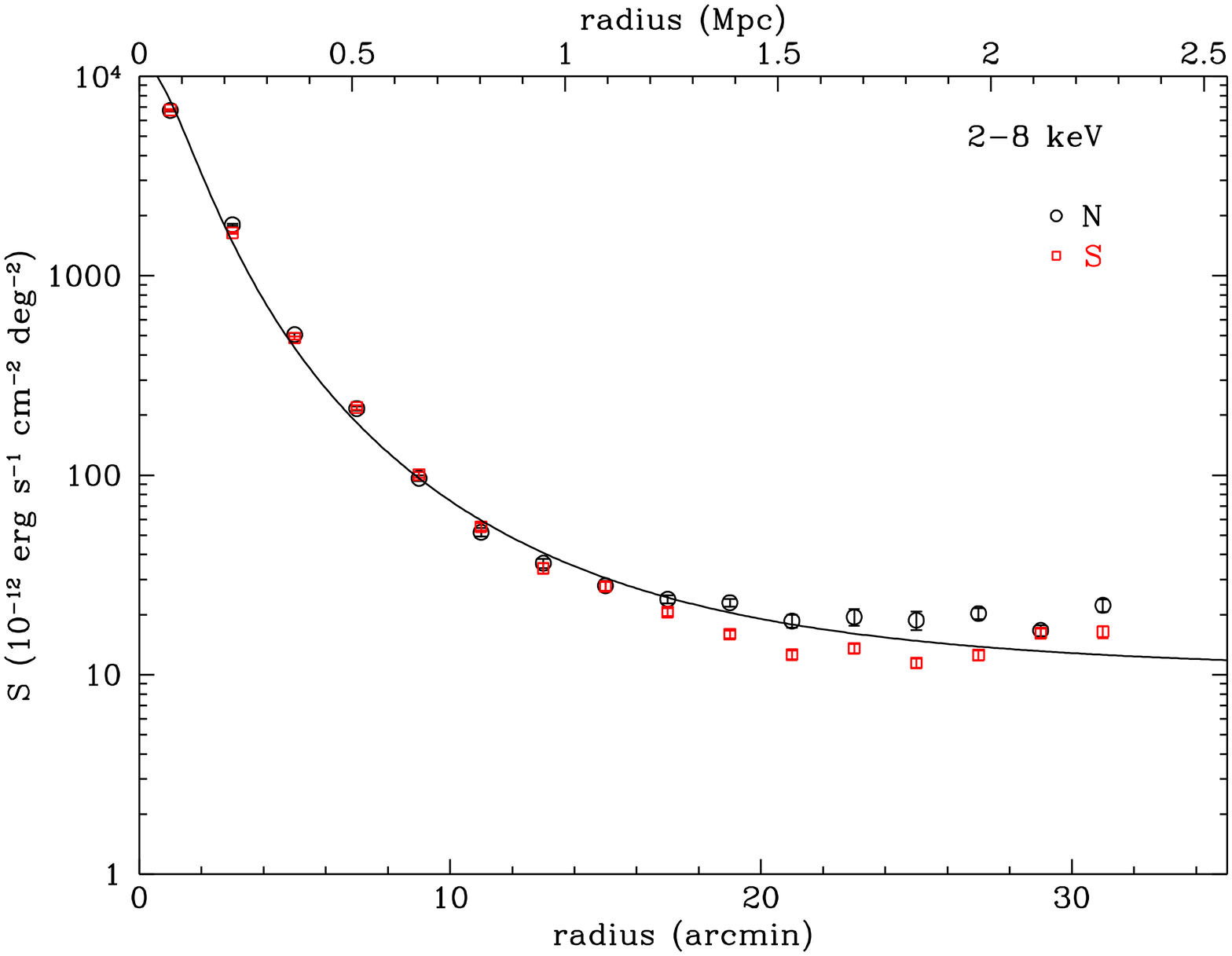}}
\caption{Surface brightness profiles northward (black circles) and
southward (red squares) from the cluster center in the 0.5--2 keV (upper
panel) and 2--8 keV (lower panel) bands.  The solid curves show the
best-fit $\beta$-models to  all the points (northward and southward) in
each panel.}
\label{figsb}
\end{figure*}

Radial surface brightness profiles were extracted from exposure-corrected
images in the 0.5--2 keV and 2--8 keV spectral bands using the Chandra
X-ray Center's CIAO/Sherpa tools.  Point sources indicated in
Figure~\ref{fig1} were excluded.  The center of the profile was determined
from a fit of a circular beta-model ($S(r) \sim (1 + x^{2})^{-(3\beta -
\frac{1}{2})}$ with $x = \frac{r}{r_{c}}$) to the (two-dimensional) 0.5--2
keV image.  A one-dimensional model consisting of a $\beta$-model plus a
constant background  was  then fit to each of the radial profiles. The
profiles and
the best-fitting models  are presented in  Figure~\ref{figsb}.  
The count-rate to flux conversions for these profiles were derived from the
SWCX-free spectral model of the emission from the background regions (see
the upper panel of Table~\ref{tab:fluxes}).
As explained in Section \ref{sexpmap}, this leads to a cluster flux
overestimate of less than 10\% (5\%) for the 0.5--2 keV (2--8 keV) band.
It is appropriate for our null hypothesis that surface brightness
variations in the A1795 outskirts are due to cosmic background
fluctuations.

The northern and southern surface brightness profiles shown in 
Figure~\ref{figsb} agree remarkably well with one another (and with a 
simple $\beta$-model with $\beta = 0.64 \pm 0.01 $) for 
$r < 1.0$ Mpc (14\arcmin) $\approx r_{2500}$.  
As expected, the cluster is symmetrical and evidently relaxed in this
region. 
At larger radii, the profiles diverge, with the surface brightness
apparently falling more rapidly than the $\beta$-model in the south, but
actually rising in the north to an apparent peak near 
$r = 1.9$ Mpc (26\arcmin) $= r_{200}$ in the 0.5--2 keV band. The surface 
brightness in this peak is about twice that in the nominal background
regions, which are at $r > 2.0$ Mpc ($r > 27.5\arcmin$).  The southern
profile actually drops $\sim 20$\% below the level in the nominal
background regions.  

\begin{figure*}
\centerline{\FigureFile(4.5in,3.375in){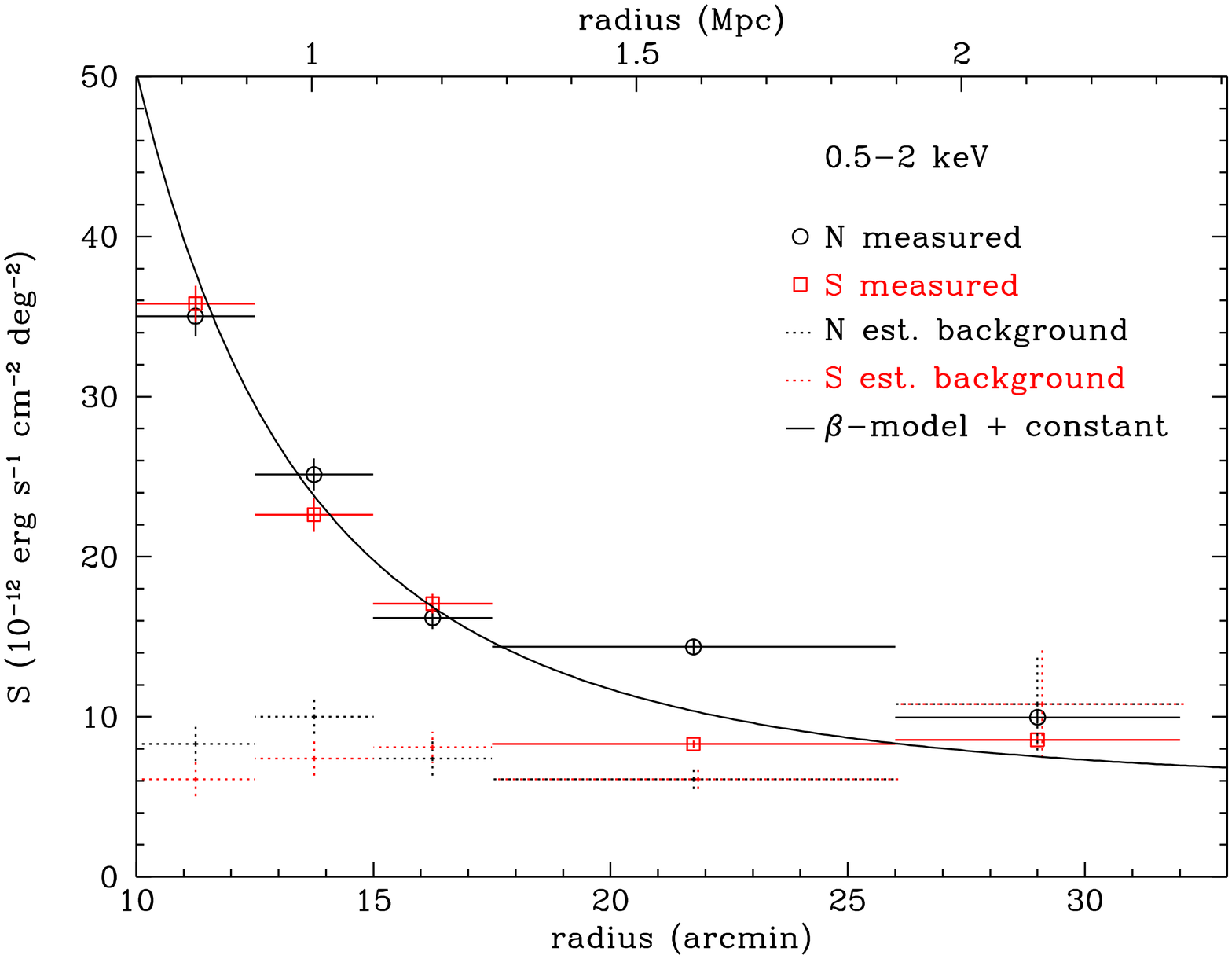}}
\centerline{\FigureFile(4.5in,3.375in){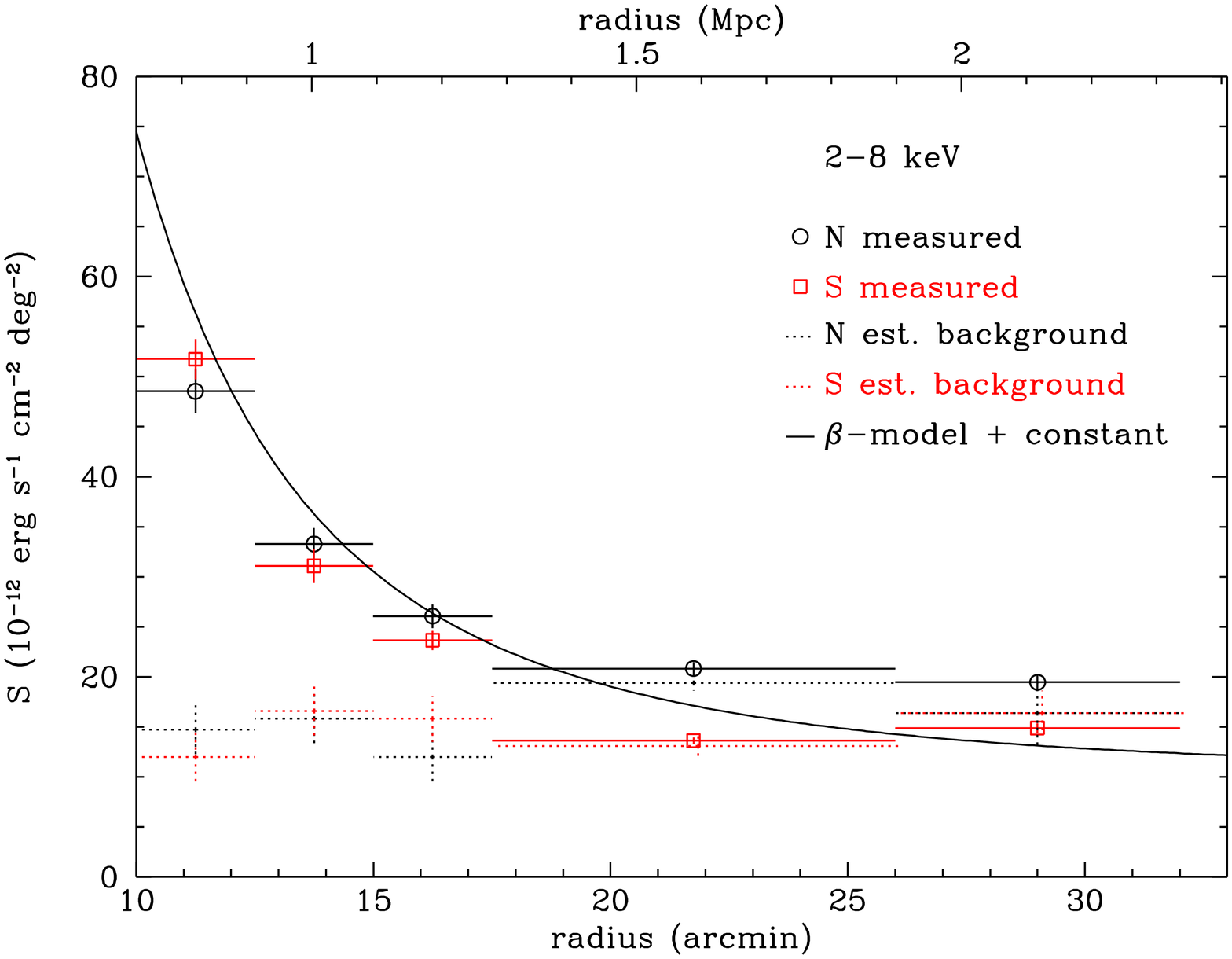}}
\caption{Surface brightness profiles at large radius in the 0.5--2 keV
(top) and 2--8 keV (bottom) bands. Profiles running northward (black
circles) and southward (red squares) from the cluster center, with
estimated corresponding background levels (dashed histograms; black to
north, red to south) are shown.  The solid curves are the best-fit
$\beta$-models assuming a spatially constant background.  The cluster
surface brightness data have been rebinned to match the spectral extraction
regions; the models are identical to those shown in Figure~\ref{figsb}.
The background estimates account for point sources detected by 
XMM-Newton
but unresolved by Suzaku, and assume no SWCX emission (see text).}
\label{figsboutskirts}
\end{figure*}

The surface brightness profiles at radii $r > 10\arcmin$ are re-plotted in
Figure~\ref{figsboutskirts}, along with  estimates  of the cosmic X-ray
background as a function of radius. To minimize the Poisson fluctuations in
the extragalactic background we have re-binned the data at $r > 18\arcmin$.
The background estimates, shown as dashed histograms in
Figure~\ref{figsboutskirts}, include three components. We assume a
spatially uniform Galactic contribution in the 0.5--2 keV band,
with flux equal to the mean thermal flux obtained from the spectral fitting
of the background regions (see Table~\ref{tab:fluxes}):
$B_{Gal} = 3.0 \pm 0.3 \times 10^{-12}$ erg s$^{-1}$ cm$^{-2}$ deg$^{-2}$
($1\,\sigma$ errors).
This estimate assumes no SWCX emission; we consider the effects of SWCX
below.  We adopt as the extragalactic background the sum of the total flux
of any point sources detected by XMM-Newton plus an estimate of the
unresolved flux derived from our adopted model of number counts
\citep{mo03}. The latter two components are listed for each spectral
analysis region in Table~\ref{bgtab}.  Error bars plotted on the background
histograms include the expected RMS Poisson fluctuations, also as listed in
Table~\ref{bgtab}. The global best-fit $\beta$-models (assuming a spatially
uniform background) are shown as solid curves in Figure~\ref{figsboutskirts}.

Figure~\ref{figsboutskirts} shows that the background model agrees well 
with the measured surface brightness in the background regions (at 
$r > 26\arcmin$), which suggests that our extragalactic background model is
reasonable.  In the region $17.5\arcmin < r < 26\arcmin$ 
($1.3 < r < 1.9$ Mpc; $r_{500} < r < r_{200}$), the observed
soft-band emission significantly exceeds the background in both the north
and the south.  The formal net soft-band cluster surface
brightnesses in this radial range are 
$B_{cl} = 8.3 \pm 0.8~(2.2 \pm 0.6) \times 10^{-12}$ erg s$^{-1}$ cm$^{-2}$ deg$^{-2}$ 
in the north (south), quoting $1\,\sigma$ errors.  
Allowing for scattered flux contributions listed in Table~\ref{tab3},
the north detection is significant at a level greater than
$8\,\sigma$; the net cluster surface brightness in the south falls
below $3\,\sigma$.

If we allow for possible SWCX contributions, the 
estimated Galactic plus geocoronal background is somewhat higher, 
but we still find significant cluster emission in the north.
In this case, the estimated Galactic plus geocoronal background 
rises to $B_{Gal+SWCX} = 4.3 \pm 0.5~(4.1 \pm 0.5) \times 10^{-12}$ erg
s$^{-1}$ cm$^{-2}$ deg$^{-2}$ in the north (south), with $1\,\sigma$
errors. 
The corresponding results for cluster flux are 
$B_{cl} = 7.0 \pm 0.8~(1.1 \pm 0.6) \times 10^{-12}$ erg s$^{-1}$
cm$^{-2}$ deg$^{-2}$ in the north (south), with $1\,\sigma$ errors.  In the
north, the signal exceeds the background by at least $6.8\,\sigma$, 
allowing for the scattered light contributions listed in Table~\ref{tab3}.
The $3\,\sigma$ upper limit on the cluster emission in the south is $B_{cl}
< 1.8 \times 10^{-12}$ erg s$^{-1}$ cm$^{-2}$ deg$^{-2}$.

We also note that formally the measured net cluster surface brightness in
the north exceeds that in the south by a factor of $6.4 \pm 3.6$ ($1\,\sigma$
error); the absolute brightness difference of 
$5.9 \pm 1.1 \times 10^{-12}$ erg s$^{-1}$ cm$^{-2}$ deg$^{-2}$ 
is significant at more than
$5.3\,\sigma$.  We conclude that Suzaku has detected soft-band cluster
emission with high confidence in a region with $r_{500} < r < r_{200}$
north of the cluster center, but not in the corresponding region in the
south.  In this radial range, the cluster is significantly brighter in the
north than in the south.

The lower panel of Figure~\ref{figsboutskirts} shows that Suzaku
detects no hard-band cluster emission at $r > 17.5\arcmin$ (1.3 Mpc) in
either north or south.  Indeed, the Suzaku surface brightness
measurements are remarkably consistent with the background estimates
derived from the XMM-Newton point-source catalog and the assumed
number-counts model for $17.5\arcmin < r < 26\arcmin$. In the hard band, the
significant north-south difference in measured Suzaku surface
brightness is attributable entirely to differences in the
background source populations in these two regions. 

Finally, comparison of the top and bottom panels of
Figure~\ref{figsboutskirts} shows that our background model implies
spatial variations in the spectral shape of the background. For example,
the background is marginally harder in the north than in the south in the
radial range $17.5\arcmin < r < 26\arcmin$, and in the north the background is
harder at $17.5\arcmin < r < 26\arcmin$ than at $r > 26\arcmin$.  We
allow for these variations in the spectral fitting used to extract the
temperature profiles discussed below. 

Simulations of cluster X-ray emission in the vicinity of  the virial
radius~\citep{ron06}  suggest that the surface brightness profile follows a
broken power law with the break at $r \approx r_{200}$. Simple power-law
fits to the A1795 surface brightness profiles at $r <  r_{200}$ are roughly
consistent with these simulations. Best-fit values for the power-law index
$\alpha$, where $S(r) \propto r^{-\alpha}$, are listed in
Table~\ref{pltab}.  The two different radial ranges shown correspond to 
$r > 0.3 r_{200}$ ($r > 7.5\arcmin$) and $0.3 r_{200} < r < 0.7 r_{200}$
($7.5\arcmin < r < 17.5\arcmin$). In
the soft band, the northern and southern profile slopes are identical,
within errors, over
the smaller radial range, but the apparent excess  emission at 
$r > 0.7 r_{200}$ ($r > 17.5\arcmin$) flattens the profile there.  In all
cases the hard-band profile slopes are steeper than the corresponding 
soft-band slopes.

\begin{table}
\caption{Power-law fits to  surface brightness profiles \label{pltab}}
\begin{center}
\begin{tabular}{l  c c }
\hline
Direction                 & \multicolumn{2}{c}{Index$^{a}$}\\
                            & 0.5--2 keV & 2--8 keV \\
\hline
\multicolumn{3}{l}{$r > 7.5\arcmin$} \\
\hline
N\hspace{20ex}         & $2.63^{+0.32}_{-0.31}$ & $3.79^{+0.44}_{-0.38}$ \\
S         & $3.43^{+0.29}_{-0.28}$ & $4.06^{+0.45}_{-0.40}$ \\
N+S       & $2.98^{+0.21}_{-0.20}$ & $3.85^{+0.31}_{-0.29}$ \\
\hline
\multicolumn{3}{l}{$7.5\arcmin < r < 17.5\arcmin$} \\
\hline
N    & $3.10^{+0.36}_{-0.36}$ & $3.51^{+0.51}_{-0.47}$ \\
S    & $3.30^{+0.33}_{-0.32}$ & $3.88^{+0.48}_{-0.44}$ \\
N+S  & $3.20^{+0.24}_{-0.24}$ & $3.68^{+0.34}_{-0.32}$ \\
\hline
\multicolumn{3}{l}{\parbox{72mm}{\footnotesize
Notes: \\
\footnotemark[$a$] Index $\alpha$ in $S(r) \propto r^{-\alpha}$, with 90\%
errors. \\
}}
\end{tabular}
\end{center}
\end{table}

The soft-band power-law index for the southern profile is quite similar to 
that reported by \citet{neu05} from stacked ROSAT profiles for a number of hot 
clusters. For example, Neumann finds $\alpha=3.79^{+0.39}_{-0.37}$ 
for $0.3 r_{200} < r  < 1.2 r_{200}$ for a set of clusters  similar to (and
including) A1795. On the other hand, the soft-band Suzaku profile
north of the cluster center is somewhat shallower than the stacked ROSAT
profile.

We have fit the rolling power-law function used by \citet{ron06} 
to our data. This function is 
$S(x) \propto x^{-\gamma(x)}$ with $x \equiv r/r_{200}$ and 
$\gamma(x) \equiv -(b_{max}x + b_{min})/(1 + x)$. At small $r$, 
$S(x) \rightarrow x^{-b_{min}}$,
while at large $r$, $S(x) \rightarrow x^{-b_{max}}$, with the transition
occurring around $x \sim 1$.  Our data are confined to $x \lesssim 1$, so
they do not constrain $b_{max}$ well.  Therefore in our fits we fix 
$b_{max} = 5.3~(7.5)$ for the soft (hard) band, values typical of the 
simulated clusters \citep{ron06}. With these constraints we find $b_{min} =
2.6$--3.5, consistent with the simulations, for both hard-band profiles and
for the soft-band profile in the south. Again because of the excess
emission at $r > 17.5\arcmin$ in the  north, however, we find $b_{min}
\approx 2.2$ for the soft-band profile in this direction; this is flatter
than the simulated clusters.

We note  that \citet{ron06} filter the emission from  their simulated
clusters before extracting surface-brightness and temperature profiles.
The aim of  this filtering is to remove high-density, low-temperature
features which are presumed to be too faint to detect, or perhaps even
non-physical artifacts of the simulations.  We note that profiles of
unfiltered, individual simulated clusters (see, for example, their Figure
2) show significant departures from the filtered soft-band surface
brightness profile, especially at  $ r > 0.5 r_{200}$.  The magnitude of
these deviations is at least as great as the north-south difference we see
in Abell 1795.  Similar fluctuations are not seen in the hard-band profiles
of the simulated clusters.  We conclude that the north-south surface
brightness difference we observe is similar to features present in these 
(unfiltered) simulated clusters.

\subsection{Temperature, Density and Entropy  Profiles}
\label{sec:tprof}

\begin{figure}
\centerline{\FigureFile(\linewidth,\linewidth){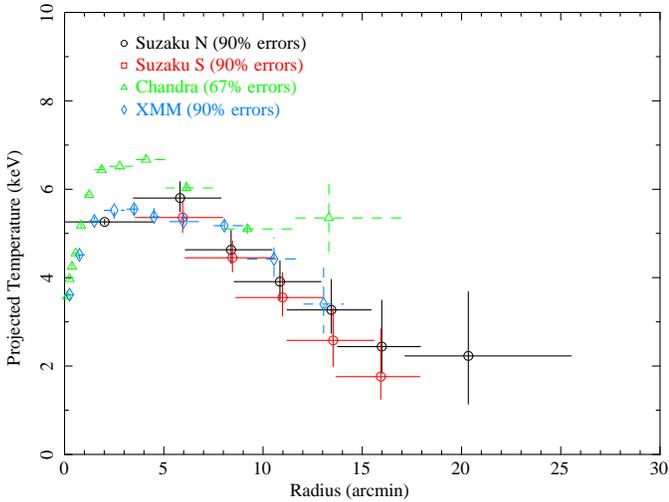}}
\caption{Projected temperature profiles running northward (black circles)
and southward (red squares) from the cluster center. Data from 
Chandra (green triangles; \cite{vik06}) and XMM-Newton (blue
diamonds; \cite{sno08}; Snowden, private communication) are also shown.  
The radius values for the Suzaku data reflect the average
originating radius of photons detected in each extraction region, as
described in Section \ref{sec:tprof}.  Likewise, the horizontal 
Suzaku error bars indicate 90\% confidence for originating location
of detected photons.  The error bars for other missions indicate the radial
range of the bin.}
\label{figtp}
\end{figure}

The temperature profiles to the north and south of the cluster are shown in 
Figure~\ref{figtp}, along with results from Chandra and 
XMM-Newton.  We are able to measure the cluster temperature  past
the limit of XMM-Newton and Chandra data (about 1.1 Mpc or 15\arcmin) 
and find that the temperature continues to decline to the largest
radius at which emission is detectable.  At the outermost radial bin,
spanning $1.3 < r < 1.9$ Mpc, the temperature is about one-third of
the peak cluster temperature observed by Suzaku. 

The radial values and 90\% error bars shown in Figure~\ref{figtp} are
estimated from the {\tt xissimarfgen} ray-tracing simulations described in
Section \ref{sec:model}.  These radial bins more accurately reflect the
real locations of photons originating from the cluster, and we use these
values for the profile fitting described in the remainder of this section.
Such corrections are less straightforward for the surface brightness
profiles shown in Figures \ref{figsb} and \ref{figsboutskirts}, since the
emission in that case is from a combination of sources (i.e., cluster and
cosmic background) that have different spatial distributions.

The Suzaku data do not resolve the cool core at $r < 2\arcmin$, but
at larger radii we find good agreement between the Suzaku and 
XMM-Newton in the radial range covered by both observatories. Note that
here we  have used the most recent analysis of the XMM-Newton data by
Snowden (private communication), which uses the latest EPIC-pn calibrations
and provides somewhat lower temperatures than those reported earlier
\citep{pif05, sno08}.  The Chandra temperatures reported by
\citet{vik06}, however, are generally higher than the Suzaku (and
XMM-Newton) temperatures outside the cool core, especially at $r <
7.5\arcmin$.  The reason for this discrepancy is not entirely clear. In
principle, scattering by the telescope of photons from the bright, cool
core over the central few arcminutes would tend to reduce the apparent
temperature.  One might expect the scattering to be more significant for
Suzaku, and, to a lesser extent,  for XMM-Newton than for 
Chandra, given the latter's much sharper point response function. If this
effect were important at radii as large as $r=7.5\arcmin$, however, it
would be surprising if Suzaku and XMM-Newton data agreed as well as
they do, given their very different point response characteristics.  A more
plausible explanation may lie in the known discrepancy between 
Chandra and XMM-Newton temperatures for relatively hot ($kT > 4$ keV)
clusters.\footnote{L. David, 2007, {\tt
http://cxc.harvard.edu/ccw/proceedings/\\07\_proc/presentations/david/}} 
Chandra reports temperatures that are systematically higher than those
from XMM-Newton for such clusters, and the magnitude of the discrepancy at
$kT_{XMM} \sim5$ keV is very close to that shown in Figure~\ref{figtp} at
$r = 5\arcmin$. The discrepancy is smaller at lower temperatures.
Uncertainties in the thickness of a thin ($\sim 10$--20 \AA) contamination
layer on the Chandra mirrors likely account for this discrepancy.  We
note that this effect is unlikely to explain the discrepancy at $r >
12\arcmin$, where $kT_{XMM} \approx 4$ keV, a temperature at which this
calibration error is thought be quite small.  We also note that
the azimuthal range covered by the Chandra observations
~\citep{vik05} differs from that we observed with Suzaku at these
large radii.  In principle, therefore,  the discrepancy between 
Chandra and Suzaku results at this large radius could reflect
azimuthal temperature variations in the cluster. In any event, we are
principally concerned here with the cluster temperature profile at $r >
5\arcmin$. Given the good agreement between the Suzaku and 
XMM-Newton temperature profiles in this radial range, we base our
subsequent analysis on the Suzaku temperatures shown in
Figure~\ref{figtp}.

Fitting a power law (assuming $T \propto r^{-\gamma}$) 
to the projected temperature at $r > 7.5\arcmin$ 
yields consistent slopes in the north and the south, and 
$\gamma=0.9\pm 0.3$ for a joint fit to all the data in this radial range. 
This is similar to the value
$\gamma=0.98 \pm 0.07$ reported by \citet{geo08} from Suzaku
observations of the outskirts of the cluster around PKS 0745-191. 

We deprojected the temperature profiles to the north and south of the 
cluster separately  using the {\tt smaug} model in {\tt XSPEC}. 
The {\tt smaug} algorithm \citep{Pizzolatoetal2003} fits analytic
functions for the deprojected temperature, density, and abundance
profiles, and compares the reprojected spectra to 
the data.  We fit spectra in the radial range  5--17.5$\arcmin$ (0.37--1.28
Mpc), although the profiles were extended to 2 Mpc for reprojection.  
Since the emission north of $17.5\arcmin$  is evidently not part of
a spherically symmetric component, we excluded it from the deprojection. We
assumed profiles of the form $f = f_0 [1+(r/r_c)^2]^{-\epsilon}$, where
$f_0$, $r_c$ and $\epsilon$ were allowed to vary independently for the
temperature and density profiles, while $f_0$ and $\epsilon$ were allowed to
vary for the abundance profile.  The abundance profile $r_c$ parameter was
frozen at 10 kpc, essentially producing a power-law dependence over the
fitting region.  We assumed a {\tt mekal} model for the plasma emission,
and included Galactic and extragalactic components identical to those
described in Section 2.  Parameter errors were estimated using  MCMC
techniques identical to those described in Section \ref{swcx}.

\begin{figure}
\centerline{\FigureFile(\linewidth,\linewidth){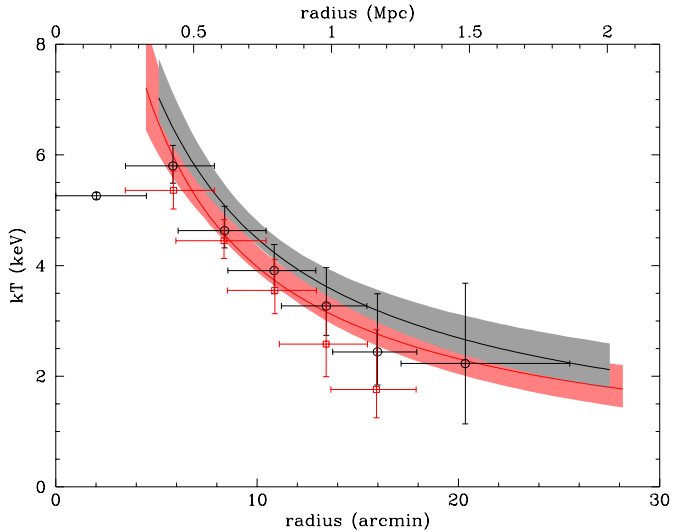}}
\caption{Deprojected temperature profiles (solid lines) and 90\%
confidence intervals (shaded regions) plotted with the observed projected
temperatures from Figure \ref{figtp} (points).  Results for the
north are in black, those for the south are in red.  Only the annuli
between 5--17.5$\arcmin$ were included in the deprojection.}
\label{figdepro}
\end{figure}

The best-fit deprojected power-law temperature profiles are shown as
continuous curves in Figure~\ref{figdepro}, along with the measured
(projected) temperatures. The shaded regions surrounding the curves
indicate 90\% confidence intervals. The  best-fit model parameters for
deprojected  temperature, hydrogen density, and abundance profiles are
listed in Table~\ref{dptab}.   Figure~\ref{figdepro} shows that the model
deprojected temperatures track the observed projected temperatures quite
closely.  As expected, the deprojected temperatures are slightly higher,
but, as suggested by~\citet{geo08}, the difference is relatively small,
especially at large radius. Note that formally the fit quality is
acceptable in the south but marginal in the north (see Table~\ref{dptab}).

\begin{table*}
\caption{Best-fit Parameters for Deprojected Profiles$^{a}$} \label{dptab}
\begin{center}
\begin{tabular}{llccc}
\hline
     &               & north & south & average \\
\hline
$kT$      & $f_0$ (keV)       &              12.8$^{+1.1}_{-0.9}$ &    16.2$^{+1.7}_{-1.3}$ &    14.5$^{+1.4}_{-1.1}$ \\
          & $\epsilon$        &              0.38$^{+0.12}_{-0.09}$ &  0.40$^{+0.12}_{-0.10}$ &  0.39$^{+0.12}_{-0.09}$ \\
          & $r_c$ (Mpc)       &              0.19$^{+0.10}_{-0.06}$ &  0.13$^{+0.09}_{-0.05}$ &  0.16$^{+0.10}_{-0.05}$ \\
\hline
$n_{\rm H}$ & $f_0$ (10$^{-3}$ cm$^{-3}$) &  3.03$^{+0.10}_{-0.08}$ & 3.06$^{+0.11}_{-0.09}$ & 3.04$^{+0.11}_{-0.08}$ \\
          & $\epsilon$        &              1.17$^{+0.03}_{-0.03}$ &  1.22$^{+0.03}_{-0.03}$ &  1.20$^{+0.03}_{-0.03}$ \\
          & $r_c$ (Mpc)       &              0.30$^{+0.01}_{-0.01}$ &  0.30$^{+0.01}_{-0.01}$ &  0.30$^{+0.01}_{-0.01}$ \\
\hline                                       
abund     & $f_0$ (solar)     &              4.31$^{+0.27}_{-3.02}$ &  4.40$^{+0.19}_{-2.91}$ &  4.36$^{+0.24}_{-2.97}$ \\
          & $\epsilon$        &              0.36$^{+0.05}_{-0.14}$ &  0.36$^{+0.06}_{-0.12}$ &  0.36$^{+0.06}_{-0.13}$ \\
          & $r_c$ (Mpc)$^{b}$ &              $0.01$                 & $0.01$                 & $0.01$ \\ 
\hline
$\chi^2$/dof &                & 1875.93/1686 & 1851.83/1810 & ... \\
\hline
\multicolumn{5}{l}{\parbox{110mm}{\footnotesize
\footnotemark[$a$] Profiles are of the form $f = f_0 [1+(r/r_c)^2]^{-\epsilon}$ \\
\footnotemark[$b$] The abundance core radius was fixed.
}}
\end{tabular}
\end{center}
\end{table*}

As expected from the projected temperatures (see Figure~\ref{figtp}), the
deprojected Suzaku temperatures are somewhat lower than those
inferred from Chandra data~\citep{vik06}. Moreover, our deprojected
temperature profile falls more rapidly with increasing radius than does the
Chandra profile.  For example, at $r  = 1.3$ Mpc $\approx r_{500}$,
our (average) model profile has a logarithmic slope $-d \ln T / d \ln r
\equiv \gamma = 0.77^{+0.16}_{-0.09}$ while ~\citet{vik06} find $\gamma =
0.48 \pm 0.15$. (Here we cite 68\% confidence errors.) In interpreting this
comparison one must again bear in mind that the  Suzaku and 
Chandra observations sample different ranges in azimuth around the
cluster.  Our measured slope is also steeper than those found for simulated
clusters by both \citet{hal07}, who fit a functional form with logarithmic
slope $\gamma \sim 0.4 \pm 0.2$ at $r = r_{500}$, and by \citet{ron06}, who
report $\gamma \sim 0.4$--$0.6$ at $r < r_{200}$. In the latter case the
steeper slope is associated with simulations which include radiative
cooling.

Our deprojected density profile approaches a power law at large radius, and
the results in Table~\ref{dptab} imply $-d \ln n_{H} / d \ln r \equiv
\alpha = 2.27 \pm 0.07$ at $r= 1.3$ Mpc. This is in
agreement with the density profile slope  reported
by~\citet{vik06} from Chandra data: their fitting formula is
reasonably well-approximated by a power law with index $\alpha = 2.21$ over
the radial range we consider.  Our assumed forms for density and
temperature imply that the ICM should be polytropic at large radius, with
a best-fit index of $\Gamma = 1.3^{+0.3}_{-0.2}$.  

\begin{figure}
\centerline{\FigureFile(\linewidth,\linewidth){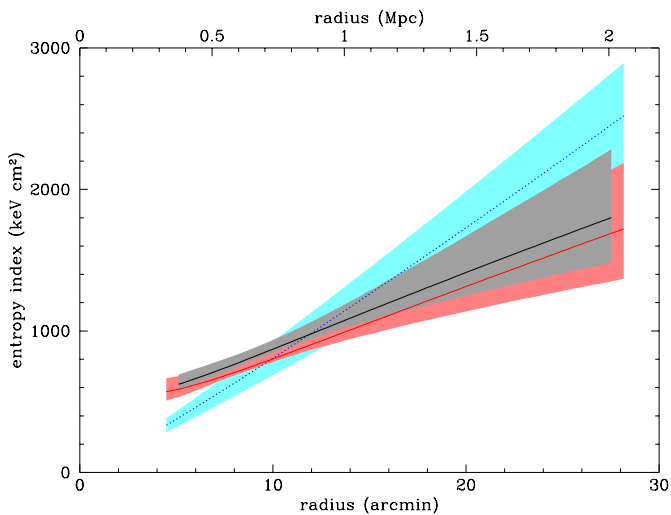}}
\caption{Entropy profiles (solid lines) and 90\% confidence intervals
(shaded regions).  Results for the north are in black, those for the
south are in red.  Plotted is the entropy index $s = kT n_e^{-2/3}$ in
keV cm$^2$.  Only annuli between 5--17.5\arcmin\ were used in the
deprojection to determine the profiles.  The blue line and shaded region
are from simulations reported by \citet{voit05} for non-radiative,
hierarchical cluster formation.}
\label{fig_ent}
\end{figure}

Finally, we have computed the implied entropy index, ($s \equiv
kT/n_{e}^{{2}/{3}}$) as a function of radius, assuming $n_{e} =
1.2\times n_{H}$.  The result is shown, with 90\%-confidence uncertainties,
in Figure~\ref{fig_ent}.  There we also show the entropy profile expected
from hierarchical structure formation derived from (non-radiative)
simulations by~\citet{voit05}; this curve increases as $s \sim r^{1.1}$.
Our average entropy profile increases more slowly, as $s \sim r^{0.74 \pm
0.20}$. 

The relatively low temperatures and entropies we observe in the outskirts
of Abell 1795 may be an indication that the plasma we detect is not in
hydrostatic equilibrium in the cluster's gravitational potential. We note
that our Suzaku observations sample two restricted sectors of the
cluster at this radius, and that data from other observatories sample
different sectors.  If the ICM is not in hydrostatic equilibrium at this
radius, one might expect azimuthal variations in ICM density and
temperature.  For example, as we noted earlier in Section~\ref{sec:sb}, the
simulations of ~\citet{ron06} produce (unfiltered) clusters with distinct
cool clumps at $r < r_{200}$. These may be the result of  relatively  cool,
low-mass substructures  falling toward the cluster center.  Moreover,
\citet{mah08} infer modest departures from hydrostatic equilibrium at radii
$r_{2500} < r < r_{500}$ from a comparison of cluster masses determined
from X-ray and weak-lensing data.  This result is at least qualitatively in
agreement with our interpretation of the temperature and entropy profiles
of Abell 1795.

\subsection{Cluster Mass}

The X-ray surface brightness and temperature profiles can be used to
estimate the cluster mass as a function of  radius if we assume that the
ICM is  hydrostatic equilibrium. In this case, following
\citet{vik06},

\begin{eqnarray}
M(<r) & = & -3.68 \times 10^{13} M_{\odot}\,\Big[\frac{T(r)}{{\rm 1\: keV}}\Big]
\Big[\frac{r}{{\rm 1\: Mpc}}\Big] \times \nonumber \\
 & & \Big[ \frac{d\log \rho_g}{d\log r} +  \frac{d\log T}{d\log r} \Big].
\label{eqn1}
\end{eqnarray}

\noindent
Here $\rho_{g}$ is the mass density of the ICM and it is assumed that the
mean molecular weight of the ICM $\mu = 0.62$. 
Chandra~\citep{vik06} and XMM-Newton~\citep{ike04,pif05}
have been used to estimate mass profiles for Abell 1795 at radii up to 
$r \lesssim 1.3$ Mpc. In this section we compare mass constraints obtained
from our Suzaku observations with these  measurements. 

Adopting the mean of the north and south profiles for the deprojected
temperature and density (see Table~\ref{dptab}), together with the mean of
the north and south temperatures measured in the final radial bin of each
deprojected profile, we use  Equation~\ref{eqn1} to estimate the cluster
mass.  We obtain $M =  4.1^{+0.5}_{-0.3} \times 10^{14} M_{\odot}$ 
within $r = 1.27$ Mpc, the value of $r_{500}$ reported by \citet{vik06}.
The 68\% confidence uncertainties quoted here are  dominated by
uncertainties in the temperature measurement.

Our mass estimate is somewhat lower than others  reported  for
Abell 1795.  \citet{vik06} find $M = 6.0 \pm 0.5 \times 10^{14} M_{\odot}$
at $r = 1.27$ Mpc, while \citet{pif05} report 
$M = 5.2 \pm 0.4 \times 10^{14} M_{\odot}$ at $r=1.16$ Mpc. Formally, these
exceed our mass estimate by $2.7\,\sigma$ and $1.7\,\sigma$, respectively.
The discrepancy is almost entirely due to the difference in temperature 
estimated at these radii.  For example, the best-fit model used by
\citet{vik06} for the deprojected temperature predicts $T = 3.8$ keV,
while we estimate $T=2.9^{+0.3}_{-0.1}$ keV at $r=1.27$ Mpc.

Our lower mass estimate  implies
$r_{500} =  1.08^{+0.07}_{-0.03}$ Mpc, somewhat smaller than the 
Chandra estimate of $r_{500} = 1.27 \pm 0.04$ Mpc~\citep{vik06}, and 
consistent with the XMM-Newton value $r_{500} = 1.16 \pm+0.05$ Mpc
~\citep{pif05}. Similarly, we find $r_{200} =  1.52^{+0.10}_{-0.07}$ Mpc, 
smaller than both the Beppo-Sax  estimate of 
$r_{200} = 2.14\pm0.46$~\citep{ett02}, and  the value ($r_{200} = 1.9$ Mpc)
we estimate from theoretical scaling relations~\citep{emn96,ea99}. Finally,
given that $r_{v} \approx r_{100}$ in our assumed 
cosmology~\citep{bn98}, our mass profile implies $r_{v} = 2.0$ Mpc.

The rapidly falling temperature profile we observe implies a rapidly
falling  mass density profile and a slowly rising integrated mass profile
(see Equation~\ref{eqn1}). A power-law approximation to our best-fit
integrated mass profile in the region $ 1 < r < 2$ Mpc yields $M(<r) = 3.7
\times 10^{14} M_{\odot} (\frac{r}{{\rm 1\: Mpc}})^{0.24}$.  This
is much flatter, for example, than the slope expected from an NFW model
with scale radius $r_{s} = 0.385$ Mpc reported by \citet{vik06}.

\section{Summary and Conclusions}

We have presented results of Suzaku observations of two distinct 
regions in the outskirts of Abell 1795.  We detect X-ray emission 
to $r = 1.3$ Mpc in both regions, and in the north trace the hot ICM 
to $r = 1.9$ Mpc $\geq r_{200}$. We find that the X-ray surface 
brightness at $1.3 < r < 1.9$ Mpc is significantly higher in the north than 
in the south. We measure the run of temperature with radius at $r > r_{2500}$ 
and find that it falls relatively rapidly ($T_{deprojected} \propto r^{-0.9}$) 
and reaches a value about one-third of its peak at the cluster
outskirts.

One possible interpretation of  our observations is that the ICM is not in
hydrostatic equilibrium in the radial range $r_{500} < r < r_{200}$, and
that in the north we are seeing infalling plasma at relatively low ($kT
\sim 2$ keV) temperature.  In support of  this picture we note that the
X-ray surface brightness is neither azimuthally symmetric nor (in the
north) falling monotonically with increasing radius. Moreover, the rapidly
falling deprojected temperature profile would require that there is
relatively  little gravitating mass at $r > 1.3$ Mpc if the ICM is
polytropic and in hydrostatic equilibrium, a result that conflicts with
extrapolations of the best-fit NFW profiles derived from high-resolution
Chandra observations~\citep{vik06}.

Simulations~\citep{ron06} predict,  and recent lensing and X-ray
observations~\citep{mah08,geo08} provide evidence for,  modest deviations
from hydrostatic equilibrium in the ICM at $r_{2500} < r < r_{200}$.  Our
observations of Abell 1795 suggest that in some cases these deviations may
be quite significant even in a cluster which appears to be relaxed at
smaller radii.  Additional observations of a variety of clusters, with more
complete azimuthal coverage than we have obtained, will improve
understanding of the state of the ICM in cluster outskirts, and therefore
of cluster masses, structure and formation mechanisms.

Finally, we note that these and other observations of the
low-surface-brightness outskirts of clusters have been made possible by
Suzaku's relatively low and stable non-X-ray background.  The
surface brightness sensitivity of our observations, 
$B_{0.5-2\,keV} > 1.8 \times 10^{-12}$ erg s$^{-1}$ cm$^{-2}$ deg$^{-2}$
($3\,\sigma$), is less than 20\% of the total cosmic background in this
spectral band.  Our sensitivity was ultimately limited by
time-variable solar wind charge exchange emission and by Poisson
variations in the extragalactic source density.  We demonstrated here
that variations in the geocoronal contribution can be modeled to this
level from simultaneous observations of cluster-free background
regions. To reduce background source density variations below this
level in the relatively small cluster regions (0.01--0.03 deg$^{2}$)
we required information from the XMM-Newton source catalog.
Future Suzaku observations of cluster outskirts would benefit
from coverage of larger ($\sim 0.1$ deg$^{2}$) solid angle in each
radial bin, especially if deep X-ray source catalogs are not
available.

\section*{Acknowledgements}
We thank Alexey Vikhlinin and Steve Snowden for providing temperature data
from Chandra and XMM-Newton, respectively, and Helen Russell 
for useful discussions about deprojection. MWB and EDM  were
supported in part by NASA grant NNG05GM92G to MIT. J.P.~Hughes acknowledges
support from NASA grant NNGG05GP87G.  J.P.~Henry acknowledges support from 
NASA grant NNG06GC04G.

\end{document}